\documentclass[twocolumn,showpacs,preprintnumbers,amssymb,nofootinbib]{revtex4}

\usepackage{graphicx,epsf,epsfig,amssymb}% Include figure files
\usepackage{bm} % Include bold math: \bm{} creates bold letters and symbols in math mode
\def\be{\begin{equation}}
\def\ee{\end{equation}}
\def\beq{\begin{eqnarray}}
\def\eeq{\end{eqnarray}}

\def\bes{\begin{eqnarray}}
\def\ees{\end{eqnarray}}

\begin{document}

\title{Perturbations and absorption cross-section of infinite-radius black rings}

\author{Vitor Cardoso}
\email{vcardoso@wugrav.wustl.edu} \affiliation{McDonnell Center for
the Space Sciences, Department of Physics, Washington University,
St.\ Louis, Missouri 63130, USA \footnote{Also at Centro de
F\'{\i}sica Computacional, Universidade de Coimbra, P-3004-516
Coimbra, Portugal}}

\author{\'Oscar J. C. Dias}
\email{odias@perimeterinstitute.ca} \affiliation{Department of
Physics, University of Waterloo, Waterloo, Ontario N2L 3G1, Canada \\
                and \\
Perimeter Institute for Theoretical Physics,  \
 Waterloo, Ontario N2L 2Y5, Canada}

\author{Shijun Yoshida}
\email{shijun@waseda.jp} \affiliation{Science and Engineering,
Waseda University, Okubo, Shinjuku, Tokyo 169-8555, Japan}
\date{\today}

\begin{abstract}
We study scalar field perturbations on the background of
non-supersymmetric black rings and of supersymmetric black rings. In
the infinite-radius limit of these geometries, we are able to
separate the wave equation, and to study wave phenomena in its
vicinities. In this limit, we show that (i) the non-supersymmetric
case is stable against scalar field perturbations, (ii) the low
energy absorption cross-section for scalar fields is equal to the
area of the event horizon in the supersymmetric case, and
proportional to it in the non-supersymmetric situation.
\end{abstract}

 \maketitle
\newpage
%%%%%%%%%%%%%%%%%%%%%%%%%%%%%%%%%%%%%%%%%%%%%%%%%%%%%%%%%%%%%%%%%%%%%%%%%
\section{Introduction}
%%%%%%%%%%%%%%%%%%%%%%%%%%%%%%%%%%%%%%%%%%%%%%%%%%%%%%%%%%%%%%%

In four dimensions, asymptotically flat black hole spacetimes have
two universal features: their horizons have spherical topology and
their properties are uniquely fixed by their conserved charges.
This is no longer true for black holes in (asymptotically flat)
higher dimensions. For instance, in five dimensions, the Einstein
theory allows not only the existence of Kerr-like black holes with
topology $S^3$, the Myers-Perry black hole \cite{myersperry}, but
also black holes with topology $S^1\times S^2$, found by Emparan
and Reall and dubbed as rotating black rings \cite{EmpReall}. In
five dimensions, a black hole can rotate along two distinct
planes. The rotating black ring of \cite{EmpReall} has angular
momentum only along $S^1$. Setting one of the angular momenta
equal to zero in the Myers-Perry black hole, there is an upper
bound for the ratio between the angular momentum $J$ and the mass
$M$ of the black hole, $J^2/M^3\leq 32/(27\pi)$, while for the
black ring there is a lower bound in the above ratio, $J^2/M^3\geq
1/\pi$, i.e., there is a minimum rotation that is required in
order to prevent the black ring from collapsing. Now, what is
quite remarkable is that for $1/\pi\leq J^2/M^3\leq 32/(27\pi)$
there are spherical black holes and black rings with the same
values of $M$ and $J$. This clearly shows that the uniqueness
theorems of four dimensions cannot be extended to non-static black
holes in five dimensions. One of the parameters that characterizes
the black ring is its radius, $R$. When this radius goes to zero
the black ring reduces to the Myers-Perry black hole, while in the
infinite-radius limit it yields a boosted black string. The black
ring solution can be extended in order to include electric charge
\cite{Elv,ElvEmp}, as well as magnetic dipole charge \cite{Emp}
(the existence of this last solution was suggested in
\cite{Reall}). The most general known seven-parameter family of
black ring solutions, that includes the above solutions as special
cases, was presented in \cite{ElvEmpFig}. It is characterized by
three conserved charges, three dipole charges, two unequal angular
momenta, and a parameter that measures the deviation from the
supersymmetric configuration. It is generally believed that these
non-supersymmetric solutions are classically unstable. The process
of Penrose extraction was analyzed in the black ring background
\cite{NozMaeda}, and a ultrarelativistic boost of the black ring
was considered in \cite{ortaggio}. Recently, a different black
ring solution with angular momentum along one of the axis of $S^2$
was found in \cite{MishIguchi}, but it has conical pathologies.

Bena and Kraus conjectured that supersymmetric black rings should
also exist \cite{BenaKraus1,Bena}. Indeed, the first example of a
supersymmetric black ring that is a solution of five-dimensional
minimal supergravity was found by Elvang, Emparan, Mateos and Reall
\cite{ElvEmpMatReall1}, using the method developed in
\cite{GauntGutHullPakisReall}, and soon after, Gauntlett and
Gutowski found a solution that describes a set of concentric
supersymmetric black rings\cite{GauntGut1}. These discoveries
triggered the research of more general supersymmetric black ring
solutions \cite{BenaWarner}-\cite{Marolf}. These studies, together
with \cite{ElvEmpFig}, progressively confirmed that upon oxidation
to higher dimensions supersymmetric black rings become equivalent to
another class of solutions known as supertubes
\cite{MatTown}-\cite{BakOhtaShe}, as first suggested in
\cite{Elv,ElvEmp}. It is now well established that the more general
supersymmetric black ring can be obtained by dimensional reduction
of an eleven dimensional supertube solution of M-theory with three
independent charges and three independent dipole charges, that
consists of three orthogonal M2-branes that carry the conserved
charges and three stacks of M5-branes whose number gives the three
dipole charges. This solution can be Kaluza-Klein reduced to a
solution of type IIA supergravity and then dualized to a type IIB
supergravity solution. The solution then describes a three charge
D1-D5-P system with D1, D5 and Kaluza-Klein monopole dipoles. So,
the most general supertube is specified by three charges, three
dipole moments and by the radius of the black ring. It reduces to
the simplest supersymmetric black ring of \cite{ElvEmpMatReall1} in
the case of three equal charges and three equal dipoles; in the
zero-radius limit it reduces to the BMPV supersymmetric black hole
solution with spherical topology \cite{BMPV}; and in the
infinite-radius limit it yields the supersymmetric black string
solution found earlier in \cite{Bena}. The supersymmetric black
rings are expected to be stable. It has been also conjectured that
horizonless three-charge supertubes might account for the
microstates of supersymmetric five dimensional black holes
\cite{MathSaxSriv}-\cite{Mathur}. A statistical counting of the
Bekenstein-Hawking entropy of supersymmetric rings was provided in
\cite{CyrierGuicaMatStrom,BenaKraus3}, and recently further progress
on the understand of entropy properties of the black ring system has
been achieved \cite{Larsen,BWentr,BGL}. For recent detailed reviews
of the black ring system see, e.g., \cite{ElvEmpMatReall1,Gaunt}.

In this paper we will consider scalar wave perturbations both in
non-supersymmetric (Sec. \ref{sec:non-supersym}) and
supersymmetric black ring (Sec. \ref{sec:supersym}) backgrounds.
It seems impossible to separate the scalar wave equation in the
black ring (supersymmetric or not) background. Nevertheless, in
the infinite-radius limit of the non-supersymmetric black ring,
which yields a boosted black string, the wave equation does
separate (Sec. \ref{sec:Weq black string}). We will then compute
the absorption cross-section of a scalar wave that impinges on a
boosted black string (Sec. \ref{sec:Absorption}). We will also
show that this background is stable against small scalar
perturbations (Sec. \ref{sec:stability}). Finally, we will also
show that in the infinite-radius limit of the supersymmetric black
ring with three equal charges and three equal dipole moments,
which yields a supersymmetric black string, the wave equation can
be separated (Sec. \ref{sec:separation super}). We will then
compute the absorption cross-section of a scalar wave that
impinges on a supersymmetric black string (Sec.
\ref{sec:absorption supersym}).

%%%%%%%%%%%%%%%%%%%%%%%%%%%%%%%%%%%%%%%%%%%%%%%%%%%%%%%%%%%%%%%%%%%%%%%%%%%%%%%%%%%%%%%%%
\section{\label{sec:non-supersym}Scalar perturbations in the black ring and in the boosted black string}
%%%%%%%%%%%%%%%%%%%%%%%%%%%%%%%%%%%%%%%%%%%%%%%%%%%%%%%%%%%%%%%%%%%%%%%%%%%%%%%%%%%%%%%%%

%%%%%%%%%%%%%%%%%%%%%%%%%%%%%%%%%%%%%%%%%%%%%%%%%%%%%%%%%%%%%%%%%%%%%%%%%%%%%%%%%%%%%%%%%
\subsection{The black ring}
%%%%%%%%%%%%%%%%%%%%%%%%%%%%%%%%%%%%%%%%%%%%%%%%%%%%%%%%%%%%%%%%%%%%%%%%%%%%%%%%%%%%%%%%%
The five-dimensional rotating black ring found by Emparan and
Reall \cite{EmpReall}, is (here we write the metric in the form
displayed in \cite{ElvEmp} after using the results of
\cite{HongTeo})
 \beq \hspace{-1cm}ds^2&=&-\frac{F(x)}{F(y)}\left
(dt+R\sqrt{\lambda \nu}(1+y)
d\psi \right )^2 \nonumber \\
& & + \frac{R^2}{(x-y)^2} {\biggl [} - F(x)\left ( G(y)d\psi
^2+\frac{F(y)}{G(y)}dy^2\right )
\nonumber \\
& & \hspace{1cm} + F(y)^2\left
(\frac{dx^2}{G(x)}+\frac{G(x)}{F(x)}d\phi ^2\right ){\biggl ]},
\label{metric}
 \eeq
 where
\begin{eqnarray}
F(\xi) = 1-\lambda \xi\,,\qquad  G(\xi) = (1-\xi ^2)(1-\nu \xi)\,.
\end{eqnarray}
The coordinate $x$ varies between $-1\leq x \leq 1$. In order to
avoid conical singularities the period of the angular coordinates
$\phi$ and $\psi$ must be given by
 \be \Delta \phi=\Delta
\psi=\frac{2\pi \sqrt{1+\lambda}}{1+\nu}\,, \label{period}
 \ee
and the parameters $\lambda$ and $\nu$ must satisfy the relation
\be \lambda=\frac{2\nu}{1+\nu ^2}\,, \qquad  0<\nu <1\,.
\label{equil cond}
 \ee
  This condition guarantees that the rotation
of the ring balances the gravitational self-attraction of the
ring.

The black ring has a curvature singularity at $y=1/\lambda$. The
regular event horizon is at $y=1/\nu$. The ergosphere is located at
$y= \pm \infty$ (these two points are identified). The solution is
asymptotically flat with the spatial infinity being located at
$x=-1$ and $y=-1$.

%%%%%%%%%%%%%%%%%%%%%%%%%%%%%%%%%%%%%%%%%%%%%%%%%%%%%%%%%%%%%%
\subsection{The wave equation of the black ring}
%%%%%%%%%%%%%%%%%%%%%%%%%%%%%%%%%%%%%%%%%%%%%%%%%%%%%%%%%%%%%%
The evolution of a scalar field $\Phi$ is governed by the curved
space Klein-Gordon equation
 \be
  \frac{\partial}{\partial x^{\mu}}
\left(\sqrt{-g}\,g^{\mu \nu}\frac{\partial}{\partial x^{\nu}}\Phi
\right)=0\,, \label{klein}
 \ee
 where $g$ is the determinant of the
metric (\ref{metric}). Following the proposal of
\cite{hawkingross,prestidge} for the C-metric\footnote{The
motivation to try the ansatz used by \cite{hawkingross,prestidge}
in the C-metric solution comes from the fact that the black ring
can be constructed by Wick-rotating the electrically charged
Kaluza-Klein C-metric \cite{EmpReall,ChambEmpReall}. For a recent
discussion on the C-metric see, e.g., \cite{DiasLemos}.}, let us
try the following wavefunction ansatz
 \be
  \Phi=(x-y)\Psi(x,y)
e^{-{\rm i}({\omega t}-m_{\phi}\phi-m_{\psi}\psi)} \,.
\label{ansatz}
 \ee
  For an equilibrium black ring the period of
$\phi$ and $\psi$ must be given by (\ref{period}). Consequently,
$m_{\phi}$ and $m_{\psi}$ must have the general form
\be m_{\phi}\,,m_{\psi}=\frac{1+\nu}{\sqrt{1+\lambda}}\,n\,,
\label{quantm} \ee
with $n$ an integer number. Using the ansatz (\ref{ansatz}) we get
\beq 0&=&\frac{\partial}{\partial x}\left
(F(x)G(x)\frac{\partial}{\partial x} \Psi \right )-
\frac{\partial}{\partial y}\left (F(y)G(y)\frac{\partial}{\partial
y} \Psi \right ) \nonumber \\
& & +\Psi {\biggl [} \frac{F(y)^2}{G(y)}\left (m_{\psi}
+R\sqrt{\lambda
\nu} \omega (1+y)\right ) ^2 \nonumber \\
& & \qquad + \frac{R^2F(y)^3\omega ^2}{(x-y)^2}-\frac{m_{\phi}^2
F(x)^2}{G(x)}{\biggr ]}
\nonumber \\
& &+\Psi {\biggl [} \frac{(F(x)G(x))'+(F(y)G(y))'}{x-y} \nonumber \\
& & \qquad +2\frac{F(y)G(y)-F(x)G(x)}{(x-y)^2} {\biggr ]}\,.
\label{kg}
 \eeq
 Using the relation
\beq & &
\frac{(F(x)G(x))'+(F(y)G(y))'}{x-y}+2\frac{F(y)G(y)-F(x)G(x)}{(x-y)^2}\nonumber
\\
& &= \nu (x-y)-2\lambda \nu (x^2-y^2) \label{kg relation}
 \eeq
in (\ref{kg}) we have
 \beq
 & & 0= \frac{\partial}{\partial x}\left
(F(x)G(x)\frac{\partial}{\partial x} \Psi \right )-
\frac{\partial}{\partial y}\left (F(y)G(y)\frac{\partial}{\partial
y} \Psi \right )+\nonumber \\
& & \Psi \left [ \frac{F(y)^2}{G(y)}\left (m_{\psi}
+R\sqrt{\lambda \nu} \omega (1+y)\right ) ^2
+\frac{R^2F(y)^3\omega ^2}{(x-y)^2}\right]+
\nonumber \\
& & \Psi \left [-\frac{m_{\phi}^2 F(x)^2}{G(x)}+\nu (x-y)-2\lambda
\nu (x^2-y^2)\right ]\,.\label{kg1} \eeq
Unfortunately, the above simplification is not enough to allow the
separability of this wave equation: the
$R^2F(y)^2\omega^2/(x-y)^2$ term seems to prevent separation.
Separability is possible only for two special situations: the
first is the time-independent or static perturbation case
($\omega=0$). This case will be discussed in appendix
\ref{sec:A0}. The other case is the infinite-radius limit,
$R\rightarrow \infty$, of the black ring. This limit yields a
boosted black string and in the next subsections we will study
wave perturbations in this background.

%%%%%%%%%%%%%%%%%%%%%%%%%%%%%%%%%%%%%%%%%%%%%%%%%%%%%%%%%%%%
\subsection{The boosted black string}
%%%%%%%%%%%%%%%%%%%%%%%%%%%%%%%%%%%%%%%%%%%%%%%%%%%%%%%%%%%%
There is a special limit of the black ring, namely the
infinite-radius limit, for which it is possible to separate the
wave equation. We will do this separation in Sec.
 \ref{sec:Weq black string}, and then, in Secs. \ref{sec:Absorption}
and \ref{sec:stability}, we shall study wave propagation phenomena
in this limit of the black ring, with the hope that the main
qualitative features of propagation in this background limit are
also valid for the general black ring.

The infinite-radius limit is defined by taking \cite{ElvEmp}
\be R \rightarrow \infty\,,\quad \lambda\rightarrow 0\,,\quad {\rm
and} \quad\nu \rightarrow 0\,, \ee
in the black ring solution (\ref{metric}), and keeping $R\lambda$
and $R\nu$ constants. In particular, take
\be R\lambda=r_H \cosh ^2 \sigma\,,\quad R\nu=r_H\sinh ^2 \sigma\,,
\ee
and introduce the following coordinates
\be r=-R\frac{F(y)}{y}\,,\quad \cos \theta=x\,,\quad \varpi=R\psi\,.
\label{boostcoord} \ee
Then the black ring solution (\ref{metric}) goes over to the
so-called boosted black string solution
\be ds^2=-\bar{f}\left( dt-\frac{r_H
\sinh{2\sigma}}{2r\bar{f}}d\varpi \right)^2+\frac{f}{\bar{f}}d\varpi
^2+\frac{1}{f}dr^2+r^2d\Omega _2 ^2\,, \label{boostedbs} \ee
where $d\Omega _2^2=d\theta ^2+\sin ^2 \theta d\phi ^2$, and
\be f=1-\frac{r_H}{r}\,,\qquad \bar{f}=1-\frac{r_H \cosh ^2
\sigma}{r}\,. \ee
The name boosted black string comes from the fact that this
solution can also be constructed by applying a Lorentz boost with
boost angle $\sigma$ to the geometry: 4-dimensional Schwarzschild
$\times \mathbb{R}$ \cite{Elv}. For an equilibrium black ring that
satisfies (\ref{equil cond}), in the infinite-radius limit one has
$\lambda=2\nu$ and the boost angle is given by $\tanh
\sigma=1/\sqrt{2}$. Our results will however apply to a general
$\sigma$.

In the above limit the points $y=1/\lambda$, $y=1/\nu$, $y= \pm
\infty$ and $y=-1$ of the black ring have, respectively, a direct
correspondence to the points $r=0$, $r=r_H$, $r=r_H \cosh^2
\sigma$ and $r=\infty$ of the boosted black string. These points
represent the curvature singularity ($r=0$), the regular event
horizon ($r=r_H$), the static limit of the ergosphere ($r=r_H
\cosh^2 \sigma$), and the asymptotic infinity ($r=\infty$) of the
boosted black string.

 Three important parameters of the boosted black string are the
temperature of its horizon $T_H$, the linear velocity of the
horizon $V_H=-\frac{g_{t\varpi}}{g_{\varpi\varpi}}{\bigl
|}_{r=0}$, and the area per unit length of the horizon $A_H$,
given by  \cite{ElvEmp}
 \begin{eqnarray}
 & & T_H=(4\pi r_H \cosh \sigma )^{-1}\,, \nonumber \\
 & & V_H=- \tanh \sigma \,, \nonumber \\
 & & A_H=4\pi r_H^2 \cosh \sigma \,.
 \label{boostedbs TVA}
\end{eqnarray}
Note also that $A_H T_H=r_H$.

%%%%%%%%%%%%%%%%%%%%%%%%%%%%%%%%%%%%%%%%%%%%%%%%%%%%%%%%%%%%%%
\subsection{\label{sec:Weq black string}The wave equation
of the boosted black string}
%%%%%%%%%%%%%%%%%%%%%%%%%%%%%%%%%%%%%%%%%%%%%%%%%%%%%%%%%%%%%%

 The evolution of a scalar field $\Phi$ in the background of
(\ref{boostedbs}) is governed by the curved space Klein-Gordon
equation (\ref{klein}). Using the ansatz
\be \Phi=\Psi(r) e^{-{\rm i}(\omega t+k\varpi)}Y_{lm}(\theta,\phi)
\,, \label{ansatz2} \ee
where $Y_{lm}$ are the usual spherical harmonics, we get the
following equation for the radial wavefunction $\Psi$ (see also
\cite{NozMaeda} for a similar wave equation)
\beq & & \left [ \frac{r^2\Delta}{\bar{f}}\omega ^2-\Delta l(l+1)-
\frac{r^2}{4\bar{f}}\left (\omega r_H\sinh 2\sigma+2k r\bar{f}\right )^2\right ]\Psi \nonumber \\
& & +\Delta \partial _r (\Delta
\partial _r \Psi)=0\,,\label{boost}
\eeq
where $\Delta \equiv r^2f=r^2-r_Hr$. At this point it worth noting
that this radial wave equation for the boosted black string can also
be obtained directly from the wave equation of the black ring
(\ref{kg1}), after applying to it the coordinate change
(\ref{boostcoord}) plus the identification $\Psi \rightarrow -y
\Psi$. In fact, the coordinate substitution (\ref{boostcoord})
implies that
\be
y=-\frac{R}{r}\bar{f}^{-1}\,\,,\,\,F(y)=\bar{f}^{-1}\,\,,\,\,G(y)=-\frac{R^2f}{r^2}\bar{f}^{-3}\,.
\ee
Then, equation (\ref{kg1}) separates and one recovers
(\ref{boost}); the term preventing separation in (\ref{kg1}),
$R^2F(y)^2\omega^2/(x-y)^2$, goes over to $\omega r^2/\bar{f}$.

At a first glance it seems like there is a singularity at the zero
of $\bar{f}$, in the wave equation (\ref{boost}). This is not
however the case since the potentially dangerous terms cancel out.
Indeed it is easy to show that the sum of the two terms proportional
to $\bar{f}^{-1}$ in (\ref{boost}) yields
\be \frac{r^2\Delta}{\bar{f}}\omega ^2\!-\! \frac{r^2r_H
^2}{4\bar{f}}\omega ^2\sinh ^2 2\sigma=\omega ^2r^3 (r+r_H\sinh ^2
\sigma)\,. \ee
We will now show that in the limit of small $\omega 's$ it is
possible to find a valid solution to equation (\ref{boost}). We
will first compute the scalar absorption cross-section of this
geometry (Sec. \ref{sec:Absorption}), and we will then show that
the metric (\ref{boostedbs}) is stable to scalar field
perturbations that might develop in the vicinity of the horizon
(Sec. \ref{sec:stability}). The method we shall use here, known as
matched asymptotic expansions, has been widely used with success
for the computation of scattering cross-section of black holes
\cite{staro1}, and also for computing instabilities of massive
fields in the Kerr background \cite{detweiler,bhb,CardDiasAdS}.

We will assume that $1/\omega \gg r_H$, i.e., that the Compton
wavelength of the scalar particle is much larger than the typical
size of the black hole; we divide the space outside the event
horizon in two regions, namely, the near-region, $r-r_H \ll
1/\omega$, and the far-region, $r-r_H \gg r_H$. These two regions
have a non-zero intersection, namely the region $r_H \ll r-r_H \ll
1/\omega$. Thus, in this overlapping region we can match the
near-region and the far-region solutions to get a solution to the
problem. This allows us to study reflection coefficients and
absorption cross-sections (Sec. \ref{sec:Absorption}). When the
correct boundary conditions are imposed upon the solutions, we
shall get a defining equation for $\omega$, and the stability or
instability of the spacetime depends basically on the sign of the
imaginary component of $\omega$. We will show that the spacetime
is stable against scalar perturbations (Sec. \ref{sec:stability}).

%%%%%%%%%%%%%%%%%%%%%%%%%%%%%%%%%%%%%%%%%%%%%%%%%%%%%%%%%%%%
\subsection{\label{sec:Absorption}Absorption cross-section of the boosted black string}
%%%%%%%%%%%%%%%%%%%%%%%%%%%%%%%%%%%%%%%%%%%%%%%%%%%%%%%%%%%%

In this section we will compute the absorption cross-section and
decay rates for a massless scalar field that impinges on a boosted
black string. More specifically, we will consider an ingoing wave
that is sent from asymptotic infinity towards the black string.
During the scattering process, part of the incident flux will be
absorbed into the event horizon and the rest will be sent back to
infinity. The problem is well defined only for $\omega>k$. Indeed,
as we shall see, the system has a potential barrier of height $k$
at the asymptotic infinity. Thus, if we want to send a wave from
infinity towards the event horizon, it will have to carry a
frequency higher than the barrier.
%%%%%%%%%%%%%%%%%%%%%%%%%%%%%%%%%%%%%%%%%%%%%%%%%%%%%%%%%%%%
\subsubsection{\label{sec:BH Near region}The near region solution}
%%%%%%%%%%%%%%%%%%%%%%%%%%%%%%%%%%%%%%%%%%%%%%%%%%%%%%%%%%%%
First, let us focus on the near-region in the vicinity of the
horizon, $r-r_H \ll 1/\omega$. We assume $\omega r_H \ll 1$ and
define the new variable
\begin{eqnarray}
z=1-\frac{r_H}{r}\,.
\end{eqnarray}
 We have $\Delta=r_H^2 z /(1-z)^2$, $\Delta
\partial _r=r_H z\partial _z$, and the horizon is at $z=0$.
 Therefore, neglecting the term $\omega^2 r^2\Delta / \bar{f}$
(this is a good approximation as long as $l\neq 0$) and
remembering that $\sinh 2\sigma=2\sinh \sigma \cosh \sigma$, the
radial wave equation (\ref{boost}) is written as
\begin{eqnarray}
& &  \hspace{-1cm} z(1-z)\partial ^2 _z \Psi +(1-z)
\partial _z \Psi  \nonumber \\
& &  \hspace{+0.5cm}+ \left [ -\frac{l(l+1)}{1-z}+
\frac{1-z}{z}\Upsilon ^2\right ]\Psi =0\,,
 \label{near wave eq}
 \end{eqnarray}
where we have defined
 \begin{eqnarray}
 \Upsilon &\equiv& r_H \cosh
\sigma(\omega-k \tanh \sigma) \nonumber \\
    &\equiv&  \frac{\omega-k |V_H|}{4\pi T_H}\,,
    \label{Upsilon}
\end{eqnarray}
with $T_H$ and $V_H$ being, respectively, the temperature and the
velocity of the horizon of the boosted black string defined in
(\ref{boostedbs TVA}). Through the definition
\begin{eqnarray}
\Psi=z^{i \,\Upsilon} (1-z)^{l+1}\,F \,,
 \label{hypergeometric function}
\end{eqnarray}
the near-region radial wave equation becomes
\begin{eqnarray}
& &  \hspace{-0.5cm} z(1\!-\!z)\partial_z^2 F+ {\biggl [} (1+i\,
2\Upsilon)-\left [ 1+2(l+1)+ i\, 2\Upsilon \right ]\,z {\biggr ]}
\partial_z F \nonumber \\
& & \hspace{1.5cm}- \left [ (l+1)^2+ i \,2\Upsilon (l+1)\right
]F=0\,.
 \label{near wave hypergeometric}
\end{eqnarray}
This wave equation is a standard hypergeometric equation
\cite{abramowitz} of the form \be z(1\!-\!z)\partial_z^2
F+[c-(a+b+1)z]\partial_z F-ab F=0 \,,\ee with
\begin{eqnarray}
& & \hspace{-0.5cm} a=l+1+i\,2\Upsilon \,,  \qquad b=l+1 \,, \qquad
c=1+ i\,2\Upsilon \,. \nonumber \\
& &
 \label{hypergeometric parameters}
\end{eqnarray}
The general solution of this equation in the neighborhood of $z=0$
is $A\, z^{1-c} F(a-c+1,b-c+1,2-c,z)+B\, F(a,b,c,z)$. Using
(\ref{hypergeometric function}), one finds that the most general
solution of the near-region equation is
\begin{eqnarray}
 \hspace{-0.5cm} \Psi &=& A\, z^{-i\,\Upsilon}(1-z)^{l+1}
F(a-c+1,b-c+1,2-c,z)\nonumber \\
& & +B\,z^{i\,\Upsilon}(1-z)^{l+1} F(a,b,c,z) \,.
 \label{hypergeometric solution}
\end{eqnarray}
The first term represents an ingoing wave at the horizon $z=0$,
while the second term represents an outgoing wave at the horizon. We
are working at the classical level, so there can be no outgoing flux
across the horizon, and thus one sets $B=0$ in
 (\ref{hypergeometric solution}). One is now interested in the
large $r$, $z\rightarrow 1$, behavior of the ingoing near-region
solution. To achieve this aim one uses the $z \rightarrow 1-z$
transformation law for the hypergeometric function
\cite{abramowitz},
\begin{eqnarray}
& \hspace{-2cm} F(a\!-\!c\!+\!1,b\!-\!c\!+\!1,2\!-\!c,z)=
(1\!-\!z)^{c-a-b} \times &  \nonumber \\
& \frac{\Gamma(2-c)\Gamma(a+b-c)}{\Gamma(a-c+1)\Gamma(b-c+1)}
 \,F(1\!-\!a,1\!-\!b,c\!-\!a\!-\!b\!+\!1,1\!-\!z)+ & \nonumber \\
&  \hspace{-0.2cm}
\frac{\Gamma(2-c)\Gamma(c-a-b)}{\Gamma(1-a)\Gamma(1-b)}
 \,F(a\!-\!c\!+\!1,b\!-\!c\!+\!1,-c\!+\!a\!+\!b\!+\!1,1\!-\!z), & \nonumber \\
&
 \label{transformation law}
\end{eqnarray}
and the property $F(a,b,c,0)=1$. The large $r$ behavior of the
ingoing wave solution in the near-region is then given by
\begin{eqnarray}
\Psi &\sim& A\,\Gamma(1-i\,2\Upsilon){\biggl [}
\frac{r_H^{-l}\,\Gamma(2l+1)}{\Gamma(l+1)\Gamma(l+1-i\,2\Upsilon)}\:
r^{l}\nonumber \\
& &
 +\frac{r_H^{l+1}\,\Gamma(-2l-1)}{\Gamma(-l)\Gamma(-l-i\,2\Upsilon)}\: r^{-l-1}
{\biggr ]}.
 \label{near field large r}
\end{eqnarray}

%%%%%%%%%%%%%%%%%%%%%%%%%%%%%%%%%%%%%%%%%%%%%%%%%%%%%%%%%%%%
\subsubsection{\label{sec:BH Far region}The far region solution}
%%%%%%%%%%%%%%%%%%%%%%%%%%%%%%%%%%%%%%%%%%%%%%%%%%%%%%%%%%%%
In the far-region, $r-r_H \gg r_H$, the wave equation (\ref{boost})
reduces to
\begin{eqnarray}
& & \hspace{-1.5cm}
\partial ^2 _r (r\Psi) +{\biggl [} \omega ^2-k^2
+\frac{r_H}{r}(\omega \sinh \sigma-k \cosh \sigma)^2
\nonumber \\
  & & \hspace{2.5cm} -\frac{l(l+1)}{r^2} {\biggr ]}(r\Psi) =0
\,.
 \label{far wave eq aux}
 \end{eqnarray}
Defining
\begin{eqnarray}
\eta ^2 & \equiv & k^2-\omega ^2 \,,\nonumber \\
\rho & \equiv &
 \frac{r_H(\omega \sinh\sigma -k \cosh \sigma)^2} {2\eta} \,, \nonumber \\
\chi & = & 2 \eta r\,,
          \label{far wave parameters}
\end{eqnarray}
then equation (\ref{far wave eq aux}) becomes
\begin{eqnarray}
\partial^2_{\chi} (\chi\Psi) +\left [-\frac{1}{4}
+\frac{\rho}{\chi}-\frac{l(l+1)}{\chi^2} \right ] (\chi\Psi)=0\,.
  \label{far wave eq}
\end{eqnarray}
This is a standard Whittaker equation \cite{abramowitz},
 $\partial^2 _\chi W +\left[ -\frac{1}{4}
 + \frac{\rho}{\chi}-\frac{1-\mu^2}{\chi^2} \right ] W=0 $, with
\begin{eqnarray}
& & \hspace{-0.5cm} W=\chi R \,,  \qquad \mu=l+1/2 \,.
 \label{Whittaker parameters}
\end{eqnarray}
The most general solution of this equation is $W=
\chi^{\mu+1/2}e^{-\chi/2}[\alpha \,M(\tilde{a},\tilde{b},\chi)+\beta
U(\tilde{a},\tilde{b},\chi)]$, where $M$ and $U$ are Whittaker's
functions with $\tilde{a}=1/2+\mu-\rho$ and $\tilde{b}=1+2\mu$. In
terms of the parameters that appear in (\ref{far wave eq}) one has
\begin{eqnarray}
& & \tilde{a}= l+1-\rho \,,
\nonumber \\
 & & \tilde{b}=2l+2 \,.
 \label{Whittaker parameters 2}
\end{eqnarray}
The far-region solution of (\ref{far wave eq}) is then given by
\begin{eqnarray}
 \Psi =\alpha \, \chi^{l} e^{-\chi/2}\,M(\tilde{a},\tilde{b},\chi)
 +\beta \, \chi^{l} e^{-\chi/2}\,U(\tilde{a},\tilde{b},\chi)  \,.
 \label{far field}
\end{eqnarray}
In the absorption cross-section problem that we are dealing with
in this section, an incident wave coming from infinity is
scattered by the black string and part of it is reflected back.
The boundary condition at infinity includes both ingoing and
outgoing waves and (\ref{far field}) has these two contributions.
We will need to identify the contribution coming from the ingoing
wave and the one due to the outgoing wave at spatial infinity, in
the large $\chi$ domain of the far-region. Now, when $\chi
\rightarrow +\infty$, one has $U(\tilde{a},\tilde{b},\chi)\sim
\chi^{-\tilde{a}}$ and $M(\tilde{a},\tilde{b},\chi)\sim
\chi^{\tilde{a}-\tilde{b}} e^{\chi}
\Gamma(\tilde{b})/\Gamma(\tilde{a})$ \cite{abramowitz}, and thus
in the large $\chi=2 \eta r$ regime, the far-region solution
behaves as
\begin{eqnarray}
\Psi \sim  \alpha \, \frac{\Gamma(2l+2)}{\Gamma(l+1-\rho)}
 \chi^{-1-\rho} e^{\chi/2} +\beta  \chi^{-1+\rho} e^{-\chi/2}
  \,.
 \label{far field large r}
\end{eqnarray}
To compute the fluxes, it will be important to note that the first
term proportional to $e^{+\chi/2}$ represents an ingoing wave,
while the second term proportional to $e^{-\chi/2}$ represents an
outgoing wave\footnote{To justify this statement first note that
in the absorption cross-section problem the frequency of the
incident wave must be greater than the potential barrier term,
i.e., in (\ref{far wave parameters}) one has $\eta ^2 = k^2-\omega
^2<0$. We choose $\eta=- i \sqrt{\omega ^2-k^2}$ and thus $\chi=-
i  2 |\eta|r$. Now, from (\ref{ansatz}) one has $\Phi\propto \Psi
e^{- i \omega t}$, and thus
 $e^{\chi/2- i \omega t}=e^{- i (|\eta|r+\omega t)}$
represents an ingoing wave while the term
 $e^{-\chi/2- i \omega t}=e^{ i (|\eta|r-\omega t)}$
describes an outgoing wave.}.

To do the matching of the far and near regions, we will also need to
know the small $\chi$ behavior, $\chi \rightarrow 0$, of the
far-region solution. In this regime one has
$U(\tilde{a},\tilde{b},\chi)\sim
\chi^{1-\tilde{b}}\Gamma(\tilde{b}-1)/\Gamma(\tilde{a})$ and
$M(\tilde{a},\tilde{b},\chi)\sim 1$ \cite{abramowitz}. The small
$\eta r$ behavior of the far-region solution is then given by
\begin{eqnarray}
 \Psi \sim \alpha \, (2\eta r)^{l} + \beta \,
 \frac{\Gamma(2l+1)}{\Gamma(l+1-\rho)}(2\eta r)^{-l-1} \,.
 \label{far field small r}
\end{eqnarray}

%%%%%%%%%%%%%%%%%%%%%%%%%%%%%%%%%%%%%%%%%%%%%%%%%%%%%%%%%%%%
\subsubsection{\label{sec:Match}Matching conditions.
 Absorption cross-section}
%%%%%%%%%%%%%%%%%%%%%%%%%%%%%%%%%%%%%%%%%%%%%%%%%%%%%%%%%%%%
When $r_H \ll r-r_H \ll 1/\omega$, the near-region solution and the
far-region solution overlap, and thus one can match the large $r$
near-region solution (\ref{near field large r}) with the small $\eta
r$ far-region solution
 (\ref{far field small r}). This matching yields
\begin{eqnarray}
& & \hspace{-1cm} \alpha =
\frac{\Gamma(2l+1)\,\Gamma(1-i2\Upsilon)}{\Gamma(l+1)\,\Gamma(l+1-i2\Upsilon)}
\frac{(2\eta)^{-l}}{r_H^l}\, A \,, \nonumber \\
& & \hspace{-1cm} \beta =
\frac{\Gamma(l+1-\rho)\,\Gamma(-2l-1)\,\Gamma(1-i2\Upsilon)}{\Gamma(2l+1)\,\Gamma(-l)\,\Gamma(-l-i2\Upsilon)}
\frac{(2\eta)^{l+1}}{r_H^{-l-1}}\, A \,,
 \label{relation amplitudes}
\end{eqnarray}
where one has $\beta \ll \alpha$, since $\eta \ll 1$.

We are now in position to compute the relevant fluxes for the
absorption cross-section. The conserved flux associated to the
radial wave equation (\ref{boost}) is
\begin{eqnarray}
 \mathcal{F}= \frac{2\pi}{i} \left( \Psi^*\Delta\partial_r \Psi
 -\Psi \Delta\partial_r \Psi^* \right)
  \,.
 \label{flux}
\end{eqnarray}
The incident flux from infinity is computed using the incoming
contribution of the large $\eta r$ far-region solution
 (\ref{far field large r}), i.e.,
$\Psi_{\rm inc}=\alpha \chi^{-1-\rho} e^{\chi/2}
\Gamma(2l+2)/\Gamma(l+1-\rho)$. The complex conjugate
 $\Psi_{\rm inc}^*$ is found noting that $\eta=- i |\eta|$
 implies, from (\ref{far wave parameters}), that
 $\rho= i |\rho|$. The derivatives of the wave solution are
 calculated using $\chi=2\eta r$, and thus
 $\partial_r=2\eta\partial_{\chi}$. After taking the limit
 $r\rightarrow \infty$, the incident flux from infinity is given
 by
 \begin{eqnarray}
 \mathcal{F}_{\rm inc}= \frac{\pi}{(\omega^2-k^2)^{1/2}}\,
 \frac{[\Gamma(2l+2)]^2}{\Gamma(l+1-\rho)\Gamma(l+1+\rho)}\,|\alpha|^2
  \,.
 \label{flux inc}
\end{eqnarray}

The flux absorbed by the boosted black string, i.e., the ingoing
flux across the horizon is computed using
 (\ref{hypergeometric solution}) with $B=0$ (recall that
 $r\rightarrow r_H$ corresponds to $z\rightarrow 0$). The derivatives of
the wave solution are taken using $\Delta \partial_r= r_H z
\partial_z$. After taking the limit
 $z\rightarrow 0$ and noting that in this limit
 $F(a,b,c,z)=[F(a,b,c,z)]^*\rightarrow 1$, the flux across the horizon is given
 by
 \begin{eqnarray}
 \mathcal{F}_{\rm abs}= A_H (\omega-k |V_H|)\,|A|^2
  \,.
 \label{flux abs}
\end{eqnarray}
where the area per unit length $A_H$ of the horizon of the boosted
black string is defined in (\ref{boostedbs TVA}).

The absorption cross-section is given by $\sigma_{\rm abs}
=[\pi/(\omega^2-k^2)] \mathcal{F}_{\rm abs}/\mathcal{F}_{\rm
inc}$. The factor $\pi/(\omega^2-k^2)$ converts the partial wave
cross-sections into plane wave cross-sections. Explicitly, the
absorption cross-section is then given by
 \begin{eqnarray}
 \sigma_{\rm abs}= A_H (\omega-k |V_H|) N \,,
 \label{cross section}
\end{eqnarray}
where $N$ is
 \begin{eqnarray}
 N&=&(\omega^2-k^2)^{l-1/2} (2 A_H T_H)^{2l}
 \frac{\Gamma(l+1-\rho)\Gamma(l+1+\rho)}{[\Gamma(2l+2)]^2} \nonumber \\
 & & \times \frac{[\Gamma(l+1)]^2}{[\Gamma(2l+1)]^2}
 \left| \frac{\Gamma(l+1- i 2\Upsilon)}{\Gamma(1- i 2\Upsilon)}\right|^2
\,.
 \label{factor cross section}
\end{eqnarray}
Using the properties of the gamma function, (\ref{gamma values 1})
and (\ref{gamma values 2}), one finds that $N$ is given by
 \begin{eqnarray}
 N&=&(\omega^2-k^2)^{l-1/2}
 \frac{(2 A_H T_H)^{2l}\,(l!)^2}{[(2l!)\,(2l+1)!]^2}\frac{\pi |\rho|}{\sinh(\pi|\rho|)}
 \nonumber \\
  & & \times \prod_{\jmath=1}^l
(\jmath^2+|\rho|^2)(\jmath^2+4\Upsilon^2)\,.
 \label{factor cross section 2}
\end{eqnarray}
The absorption cross-section (\ref{cross section}) of a boosted
black string  has some features that deserve a closer look:

 \noindent (i) The factor $N$ is a well-behaved real positive
factor as long as $\omega>k$. For $\omega<k$, $N$ and $\sigma_{\rm
abs}$ are pure imaginary numbers. Physically, we can understand this
behavior by noting that the quantum number that specifies the
momentum of wave along the boosted direction provides a natural
potential height $k$ at infinity. Therefore, if we want to send a
wave from infinity towards the boosted black string, its frequency
must be higher than the height of the potential. So the absorption
cross-section problem is only well posed when $\omega>k$, and
(\ref{cross section})  is always positive. The above requirement
together with the fact that the boost velocity satisfies
$|V_H|=\tanh \sigma \leq 1$, implies that the factor
$(\omega-k|V_H|)$ in (\ref{cross section}) is always positive. This
means that the potential barrier at infinity prevents the existence
of superradiant scattering in the boosted black string background.
This peculiar behavior was first studied \cite{NozMaeda}, and sets a
significant difference between the wave scattering on a rotating
Myers-Perry black hole and on a boosted black string. Indeed, in a
Myers-Perry black hole the absorption cross-section is proportional
to a power of $\omega$ times $(\omega-k\Omega_H)$, where $\Omega_H$
is the angular velocity at the horizon. In this case there is no
potential barrier at infinity, and thus there is no lower bound for
the frequency of the wave that we can send from infinity towards the
black hole. This allows the existence of a superradiant regime: for
$\omega<k\Omega_H$, the absorption cross-section of the Myers-Perry
black hole is negative, i.e., during the scattering process in the
ergosphere, the scalar wave extracts rotational energy from the
black hole and the amplitude of the scattered wave is bigger than
the one of the incident wave.

\noindent (ii) For black holes that have horizons with spherical
topology it is a known universal result that the low energy
absorption cross-section of a massless scalar wave is given by the
area of the black hole horizon \cite{DasGibbonsMathur}. We may ask
if this is still the case for black objects with topology
$S^1\times S^2$, as is the case of the boosted black string. One
first notes that the quantum numbers $l$ and $k$ are independent,
and thus choosing $l=0$ imposes no restrictions on the value of
$k$. So, inserting $l=0$ in (\ref{cross section})\footnote{The
expression (\ref{cross section}) is not actually accurate for
$l=0$ since in (\ref{near wave eq}) we have neglected the term
$\omega^2 r^2\Delta / \bar{f}$, which is a good approximation only
if $l\neq 0$. However, a similar situation occurs with the
computations of the absorption cross-sections of spherically
symmetric black holes that have been done previously in the
literature. In this case it was also assumed that the general
expression for the absorption cross-section could be extended to
the $l=0$ case.} one concludes that the low absorption
cross-section is not given simply by the area of the horizon, it
also depends on $\omega$ and $k$.

\noindent (iii) As some trivial checks to our expression (\ref{cross
section}), we note that when we set the boost parameter $\sigma$
equal to zero one gets the cross-section of a black string (i.e.,
Schwarzchild times $\mathbb{R}$). Moreover, if we also set $k=0$, we
recover the absorption cross-section of the Schwarzchild black hole.

Given the absorption cross-section we can also compute the decay
rate,
 \begin{eqnarray}
 \Gamma_{\rm decay}\!\!&=&\!\! \frac{\sigma_{\rm abs}}{e^{\frac{\omega-k
 |V_H|}{T_H}}-1}\nonumber \\
 &=& e^{-\frac{\omega-k |V_H|}{2T_H}}(\omega^2-k^2)^{l-1/2} 2^{2l}
  \left| \Gamma(l+1- i 2\Upsilon) \right|^2
  \nonumber \\
 & &\!\! \times (A_H T_H)^{2l+1} \frac{|\Gamma(l+1+\rho)|^2}{[\Gamma(2l+2)]^2}
  \frac{[\Gamma(l+1)]^2}{[\Gamma(2l+1)]^2}
  \,.
 \label{decay rate}
\end{eqnarray}
%%%%%%%%%%%%%%%%%%%%%%%%%%%%%%%%%%%%%%%%%%%%%%%%%%%%%%%%%%%%
\subsection{\label{sec:stability}Stability study of the boosted black string}
%%%%%%%%%%%%%%%%%%%%%%%%%%%%%%%%%%%%%%%%%%%%%%%%%%%%%%%%%%%%
In this section we will show that a boosted black string is stable
against massless field perturbations. We will first do an analytical
study of the wave equation, in a certain regime, and then we will
solve it numerically. Both results agree, in the regime where the
analytical calculation holds. One might conjecture that this metric
is unstable to superradiant effects (the mechanism at play in this
instability was recently described in
\cite{cardosobraneinst,marolf,zelmisnerunruhbekenstein,bhb,CardDiasAdS}),
since it is an extended rotating black object. The massless scalar
field acquires an effective mass (due to the extra dimension) and
this could lead to superradiant amplification in the ergosphere and
consequent instability \cite{cardosobraneinst,marolf}. Nevertheless,
we will show that boosted black strings are stable, to massless
scalar field perturbations. We still expect the Gregory-Laflamme
instability \cite{gl} to be at work here, and so the geometry should
be unstable to gravitational perturbations.
%%%%%%%%%%%%%%%%%%%%%%%%%%%%%%%%%%%%%%%%%%%%%%%%%%%%%%%%%%%%%%%%%%%%%%%%%%%%%%%%%%

%%%%%%%%%%%%%%%%%%%%%%%%%%%%%%%%%%%%%%%%%%%%%%%%%%%%%%%%%%%%
\subsubsection{\label{sec:Analytical stab}Analytical study of the stability}
%%%%%%%%%%%%%%%%%%%%%%%%%%%%%%%%%%%%%%%%%%%%%%%%%%%%%%%%%%%%

%%%%%%%%%%%%%%%%%%%%%
{\it \noindent The near region solution:} \vskip 0.2cm
%%%%%
The results found in Sec. \ref{sec:BH Near region} are also valid
in the present problem. So, the near-region wave equation is given
by (\ref{near wave eq}). Requiring only ingoing flux at the
horizon $z=0$, the near-region wave solution that satisfies the
physical boundary condition is then given by
 (\ref{hypergeometric solution}) with $B=0$. For large $r$ the
near-region solution behaves as (\ref{near field large r}), that we
reproduce here again for the sake of clarity:
\begin{eqnarray}
\Psi &\sim& A\,\Gamma(1-i\,2\Upsilon){\biggl [}
\frac{r_H^{-l}\,\Gamma(2l+1)}{\Gamma(l+1)\Gamma(l+1-i\,2\Upsilon)}\:
r^{l}\nonumber \\
& &
 +\frac{r_H^{l+1}\,\Gamma(-2l-1)}{\Gamma(-l)\Gamma(-l-i\,2\Upsilon)}\: r^{-l-1}
{\biggr ]}.
 \label{near field large r REP}
\end{eqnarray}

%%%%%%%%%%%%%%%%%%%%%%%
{\it \noindent The far region solution:} \vskip 0.2cm
%%%%%%%
The far-region solution is given by (\ref{far wave eq}) subjected to
(\ref{far wave parameters}). The most general wave solution of this
far-region is described by (\ref{far field}) and (\ref{Whittaker
parameters 2}). The behavior of the far-region solution in the large
$\chi=2 \eta r$ regime is given by
 (\ref{far field large r}), where the first term
proportional to $e^{+\chi/2}$ represents an ingoing wave, while the
second term proportional to $e^{-\chi/2}$ represents an outgoing
wave.

The stability problem differs significantly from the absorption
cross-section problem (that we dealt with in Sec.
\ref{sec:Absorption}) mainly due to the nature of the boundary
conditions that are imposed in the far-region. In the stability
problem one perturbs the boosted black string outside its horizon,
which generates a wave that propagates both into the horizon and
out to infinity. Therefore, at $r\rightarrow +\infty$, our
physical system has only an outgoing wave (while the absorption
cross-section problem of Sec. \ref{sec:Absorption} also had an
ingoing wave), and thus one sets $\alpha=0$ in (\ref{far field}):
\begin{eqnarray}
 \Psi =\beta \, \chi^{l} e^{-\chi/2}\,U(\tilde{a},\tilde{b},\chi) \,.
 \label{far field REP}
\end{eqnarray}

Now, we turn our attention to the small $\chi$ behavior, $\chi
\rightarrow 2 \eta r_H$, of the far-region solution, which is
described by (\ref{far field small r}), with $\alpha=0$ as justified
in the last paragraph. Whittaker's equation also describes the
hydrogen atom system, even though one must be cautious since the
boundaries in our case are $\chi \sim 2\eta r_H$ and $\chi=\infty$,
while in the hydrogen atom the boundaries are $\chi=0$ and
$\chi=\infty$. The wave equation for the electron wavefunction
$\Psi$ in the hydrogen atom is of the type (\ref{far wave eq}), and
at spatial infinity it is given by
 (\ref{far field REP}). The inner boundary condition is
that $\Psi$ must be regular at the origin, $\chi\rightarrow 0$. For
small values of $\chi$, the solution is described by $\Psi \sim
\beta \, [\Gamma(2l+1)/\Gamma(l+1-\rho)]
 (2\eta r)^{-l-1}$ \cite{abramowitz}.
So, when $\chi\rightarrow 0$, the wavefunction $\Psi$ diverges,
$r^{-l-1}\rightarrow \infty$. In order to have a regular solution
there we must then demand that $\Gamma(l+1-\rho)\rightarrow
\infty$. This occurs when the argument of the gamma function is a
non-positive integer, $\Gamma(-n)=\infty$ with $n=0,1,2,\cdots$.
Therefore, the requirement of regularity imposes the condition
$l+1-\rho=-n$, in the hydrogen atom. Since $\rho$ is related to
$\omega$, the demanding regularity at the inner boundary amounts
to a natural selection of the allowed frequencies in the hydrogen
atom. Now, let us come back to the boosted black string
background. In the spirit of \cite{detweiler,CardDiasAdS}, we
expect that the presence of an horizon induces a small complex
imaginary part in the allowed frequencies, $\omega_i={\rm
Im}[\omega]$, that describes the slow decay of the amplitude of
the wave if $\omega_i<0$, or the slowly growing instability of the
mode if $\omega_i>0$. Now, from (\ref{far wave parameters}), one
sees that a frequency $\omega$ with a small imaginary part
corresponds to a complex $\rho$ with a small imaginary part that
we will denote by $\delta \rho \equiv {\rm Im}[\rho] $. Therefore,
guided by the hydrogen atom, we set that in the boosted black
string case one has
\begin{eqnarray}
\rho=(l+1+n)+ i \delta\rho\,,
 \label{frequency spectrum}
\end{eqnarray}
with $n$ being a non-negative integer, and $\delta \rho$ being a
small quantity. In particular, this means that onwards, the
arguments of the Whittaker's function $U(\tilde{a},\tilde{b},\chi)$
previously defined in (\ref{Whittaker parameters}) are to be
replaced by
\begin{eqnarray}
& & \hspace{-0.5cm} \tilde{a}=-n- i \delta\rho\,,  \qquad
\tilde{b}=2l+2 \,.
 \label{Whittaker parameters end}
\end{eqnarray}

What we have done so far was to use the hydrogen atom paradigm to
{\it guess} the form of the eigenfrequencies in our case. At the
end, we will verify consistency of all our assumptions. In order to
match the far-region with the near-region, we will need to find the
small $\chi$ behavior of the far-region solution (\ref{far field
REP}). The Whittaker's function $U(\tilde{a},\tilde{b},\chi)$ can be
expressed in terms of the Whittaker's function
$M(\tilde{a},\tilde{b},\chi)$ \cite{abramowitz}. Inserting this
relation on
 (\ref{far field REP}), the far-region solution
can be written as
\begin{eqnarray}
\Psi &=& \beta \, \chi^{l} e^{-\chi/2}\,\frac{\pi}{\sin(\pi
\tilde{b})} {\biggl [}
\frac{M(\tilde{a},\tilde{b},\chi)}{\Gamma(1+\tilde{a}-\tilde{b})\Gamma(\tilde{b})}
\nonumber \\
 & &
-\chi^{1-\tilde{b}}\,
\frac{M(1+\tilde{a}-\tilde{b},2-\tilde{b},\chi)}{\Gamma(\tilde{a})\Gamma(2-\tilde{b})}
{\biggr ]},
 \label{far field-small x aux}
\end{eqnarray}
with $\tilde{a}$ and $\tilde{b}$ defined in
 (\ref{Whittaker parameters end}).
Now, we want to find the small $\chi$ behavior of
 (\ref{far field-small x aux}), and to extract $\delta \rho$ from the gamma
function. This is done in
 (\ref{far small x A1})-(\ref{far small x A3}), yielding for small
$\delta \rho$ and for small $\chi$ the result
\begin{eqnarray}
 \Psi &\sim& \beta (-1)^n \frac{(2l+1+n)!}{(2l+1)!}
\, (2\eta r)^l \nonumber \\
 & & + \beta (-1)^{n+1}(2l)! \, n! \, ( i \delta \rho) \, (2 \eta r)^{-l-1}.
 \label{far field-small x}
\end{eqnarray}

%%%%%%%%%%%%%%%%%%%%%%%
{\it \noindent Matching condition:} \vskip 0.2cm
%%%%%%%

The quantity $\delta \rho$ cannot take any value. Its allowed values
are selected by requiring a match between the near-region solution
(\ref{near field large r REP}) and the far-region solution (\ref{far
field-small x}). So, the allowed values of $\delta \rho$ are those
that satisfy the matching condition
 \begin{eqnarray}
& & \hspace{-0.8cm}  - i \delta \rho \frac{(2l)!(2l+1)!n!}
 {(2l+n+1)!(2\eta )^{2l+1}}
 \nonumber \\
 & & \hspace{-0.5cm}=r_H^{2l+1}
\frac{\Gamma(l+1)}{\Gamma(2l+1)}\frac{\Gamma(-2l-1)}{\Gamma(-l)}
\frac{\Gamma(l+1- i \,2\Upsilon)}{\Gamma(-l- i \,2\Upsilon)} \,.
 \label{Matching Instab}
\end{eqnarray}
Using  (\ref{Upsilon}) and the gamma function relations (\ref{gamma
values 3}) we get
\begin{eqnarray}
\delta \rho & =& -2  \Upsilon(2\eta
r_H)^{2l+1}\frac{(2l+1+n)!}{n!}  \nonumber \\
& & \times \left [ \frac{l!}{(2l)!(2l+1)!}\right
]^2\prod_{\jmath=1}^l (\jmath^2+4\Upsilon ^2)\,.
 \label{delta rho}
\end{eqnarray}
For $l=0$ this would be \be \delta \rho =-4\eta r_H (n+1)
\Upsilon\,.\ee
 Condition ({\ref{frequency spectrum}}) together with
  (\ref{far wave parameters}) leads to
\beq \frac{r_H(\omega \sinh \sigma-k \cosh
\sigma)^2}{2\sqrt{k^2-\omega ^2}}=l+n+1+i \delta \rho\,,
\label{condfinal}\eeq
 with $\delta \rho$ given by (\ref{delta rho}).
This is therefore an algebraic equation for the characteristic
values of the frequency $\omega$. All these values must be
consistent with the assumptions made: $k\ll 1$ and $\omega \ll 1$.
If $\omega$ has a positive imaginary part, then the mode is
unstable: because the field has the time dependence $e^{-i\omega
t}$, a positive imaginary part for $\omega$ means the amplitude
grows exponentially as time goes by. We have performed an
extensive search for unstable modes, and didn't find any.
Condition (\ref{condfinal}) is solved by many values of
$\omega\,,k\,,\sigma$, but the only ones satisfying the
assumptions seem to be marginally stable modes with $\omega \simeq
k$, i.e., purely real frequencies. As we shall see in the next
section, these purely real modes are also found numerically.

%%%%%%%%%%%%%%%%%%%%%%%%%%%%%%%%%%%%%%%%%%%%%%%%%%%%%%%%%%%%%%%%%%%%%
\subsubsection{\label{sec:Numerical Instab}Numerical analysis}
%%%%%%%%%%%%%%%%%%%%%%%%%%%%%%%%%%%%%%%%%%%%%%%%%%%%%%%%%%%%%%%%%%%%%
The appropriate boundary conditions for a (regular) scalar field
evolving in this geometry were described in the previous section:
ingoing waves at the horizon, and outgoing waves near infinity.
These boundary conditions together with the wave equation
(\ref{boost}) form an eigenvalue problem for $\omega$: only for
certain values of $\omega$ one can satisfy both conditions
simultaneously. The characteristic frequencies that do so are
called quasinormal frequencies \cite{kokkotas}. The numerical
method used here to compute the quasinormal frequencies is
described in more detail in Appendix \ref{sec:A2}.

\begin{figure}
\centerline{ \mbox{ \psfig{figure=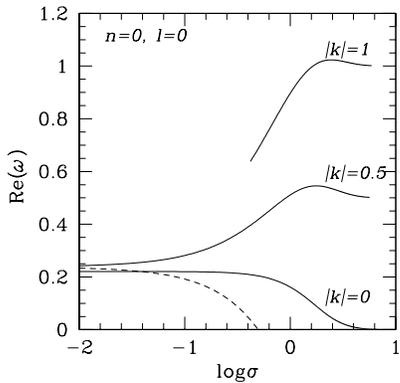,angle=0,width=6cm}}}
\caption{The real part of the fundamental mode ($n=0$) of $l=0$
perturbations, as a function of the boost parameter $\sigma$. Here
we show the characteristic frequencies for several values of $k$.
For small values of $\sigma$ and $k$, the real part of the
frequency yields ${\rm Re}(\omega) \sim 0.22 $, which is the
Schwarzschild value.} \label{fig:f1}
\end{figure}

\begin{figure}
\centerline{ \mbox{ \psfig{figure=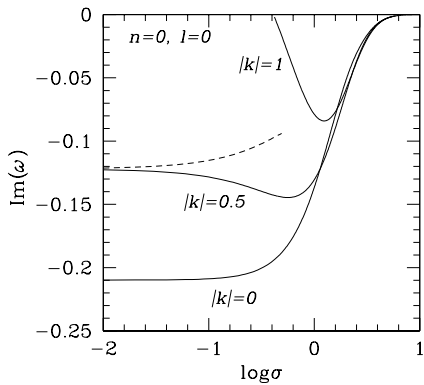,angle=0,width=6cm}}}
\caption{The imaginary part of the fundamental mode ($n=0$) of
$l=0$ perturbations, as a function of the boost parameter
$\sigma$. For small values of $\sigma$ and $k$, the imaginary part
of the frequency yields ${\rm Im}(\omega) \sim 0.21$, which is the
Schwarzschild value.} \label{fig:f2}
\end{figure}

\begin{figure}
\centerline{ \mbox{ \psfig{figure=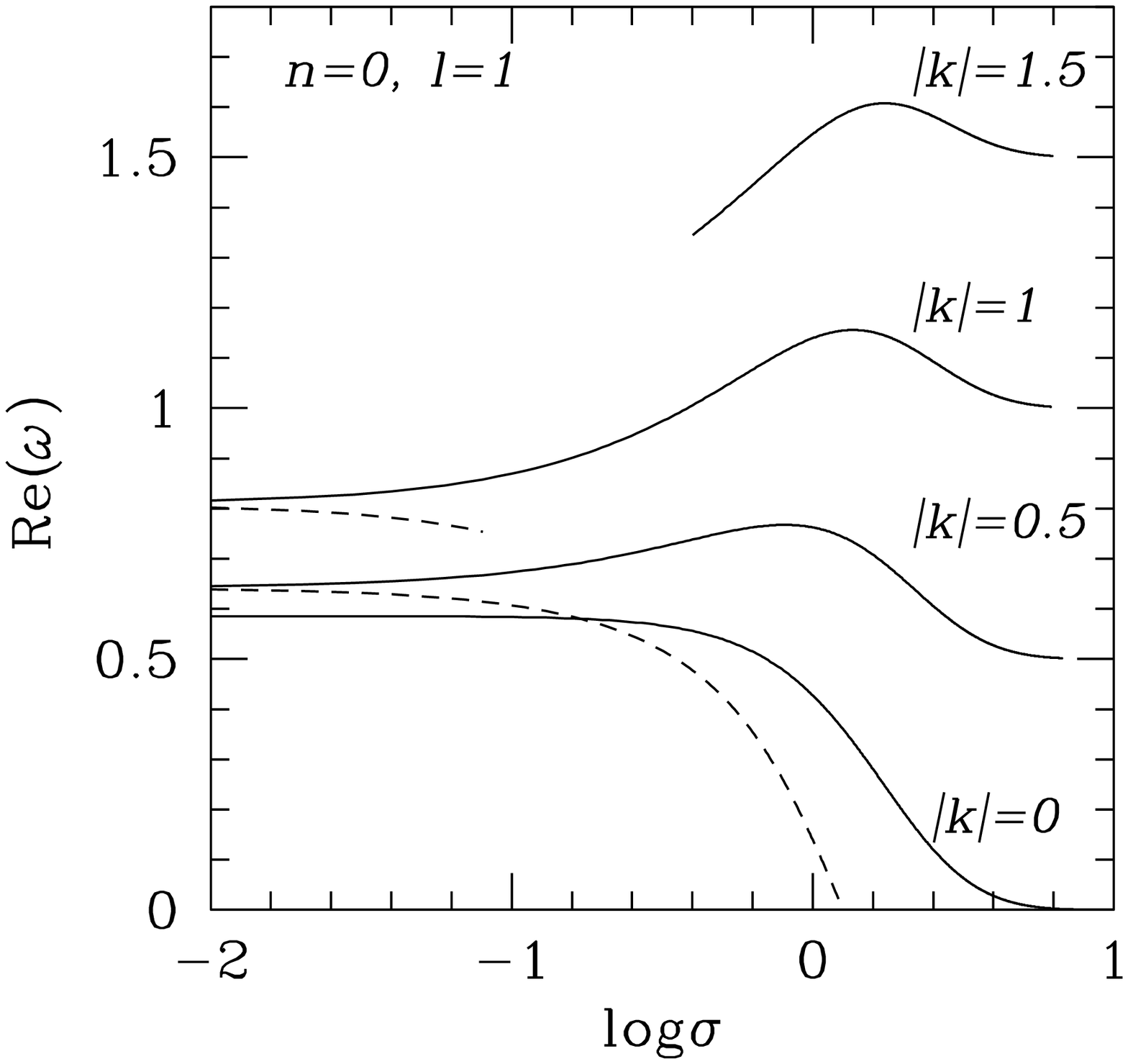,angle=0,width=5cm}
\psfig{figure=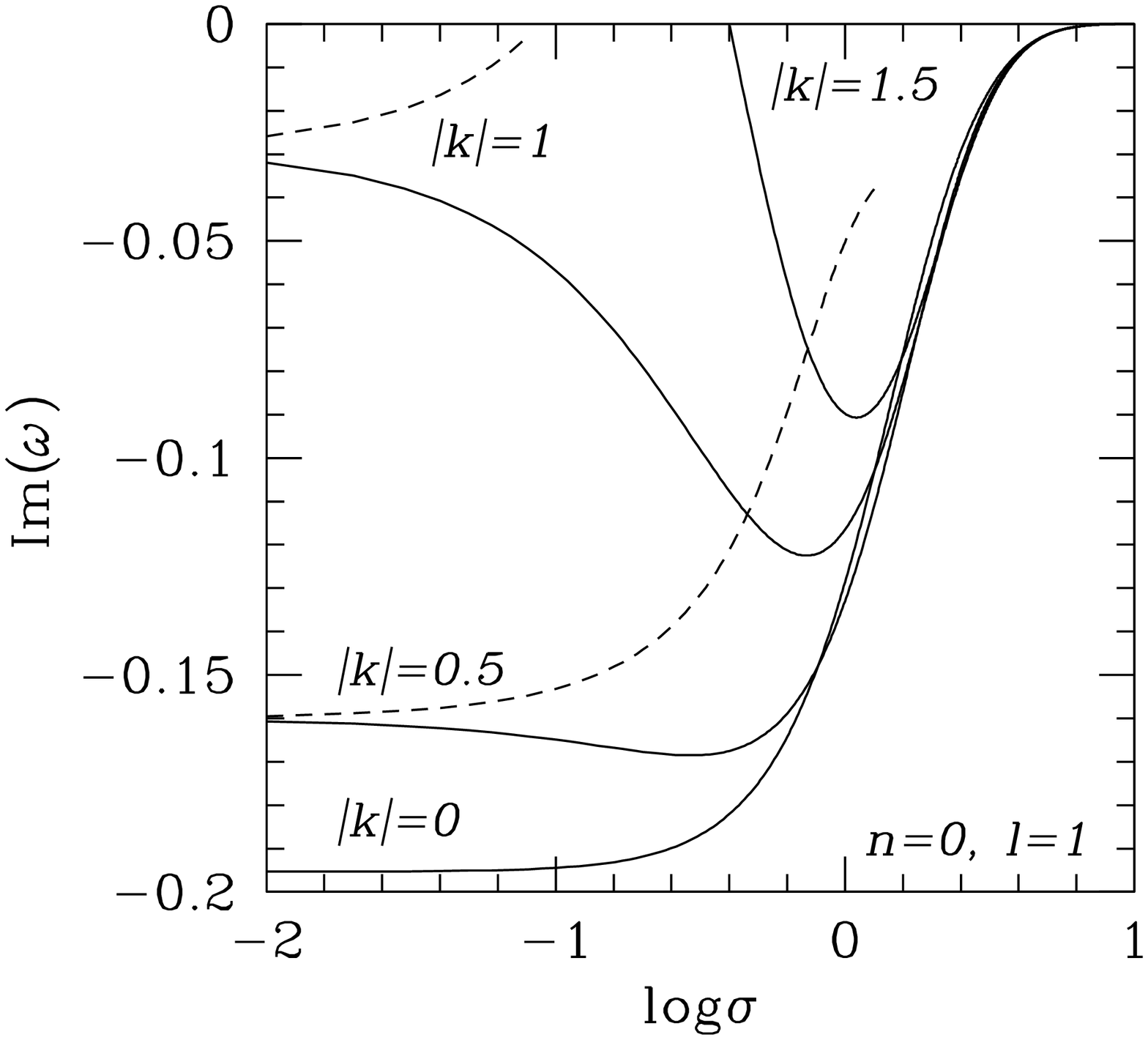,angle=0,width=5cm} }} \caption{The same as Figs
\ref{fig:f1} and \ref{fig:f2}, but for $l=1$.} \label{fig:f3}
\end{figure}

Using the numerical code, we have looked for stable and unstable
modes of the boosted black string. We found only stable modes,
which indicates that this boosted black string is stable to
scalar perturbations. The numerical results for the stable modes
are shown in Figs. \ref{fig:f1}-\ref{fig:f3}. For each
$k\,,\,l\,,\,,\sigma$, there is a large number (possibly an
infinity) of frequencies that satisfy the boundary conditions. We
order them by magnitude of imaginary part. Thus, the mode with
lowest imaginary part (in magnitude) is called the fundamental
mode and labeled with an integer $n=0$, the mode with the second
lowest imaginary part is called the first overtone $n=1$, etc.

For very small values of $\sigma$ we expect to recover the well
known Schwarzschild results \cite{iyer,willsimone}, since in this
limit the boosted black string is just a Schwarzschild black hole,
with an extra dimension. Thus, the characteristic frequencies
should be equal to the characteristic frequencies of a {\it
massive} field (with mass $k$) in the Schwarzschild spacetime.
Now, the values of these frequencies are listed in \cite{iyer} for
massless fields and are computed in \cite{willsimone} for general
massive fields. The fundamental frequency of massless scalar
fields in the Schwarzschild spacetime is \cite{iyer} $\omega r_+
\sim 0.221-i\,0.2098$ for $l=0$ and $\omega r_+ \sim 0.5858-i\,
0.1954$ for $l=1$. It is apparent from Figs.
\ref{fig:f1}-\ref{fig:f3} that for small values of $k$ and
$\sigma$, the real part and imaginary part do go to the
Schwarzschild results. For large values of the boost parameter
$\sigma$ the numerical results show that the real part of the
frequency tends to $k$, ${\rm Re}(\omega) \sim k$, while the
imaginary part goes to zero in a $k$-independent manner. This is
basically the regime found through the analytical approach in the
previous section.

%\begin{figure}
%\centerline{ \mbox{ \psfig{figure=f5.eps,angle=0,width=5cm}
%\psfig{figure=f6.eps,angle=0,width=5cm}}} \caption{Write caption.}
%\label{fig:f4}
%\end{figure}

The results for boosted black strings are in a way the same as the
``boosted results'' of unboosted black strings. We know that
unboosted black strings are stable to scalar perturbations
\cite{gregory2}, and thus one could expect boosted black strings
to be trivially stable. However, the actual situation is not as
simple, because, for example, when one compactifies after a boost
(so the solution changes globally) one gets ergosurfaces that were
not there before the boost. This gives rise to new phenomena, such
as the Penrose process, etc, that were not present in the
unboosted solution. Thus, stability of boosted black strings does
not seem to follow trivially from stability of unboosted black
strings.

%%%%%%%%%%%%%%%%%%%%%%%%%%%%%%%%%%%%%%%%%%%%%%%%%%%%%%%%%%%%%%%%%%%%%%%%%
\section{\label{sec:supersym}Scalar perturbations of the supersymmetric black ring and black string}
%%%%%%%%%%%%%%%%%%%%%%%%%%%%%%%%%%%%%%%%%%%%%%%%%%%%%%%%%%%%%%%
%%%%%%%%%%%%%%%%%%%%%%%%%%%%%%%%%%%%%%%%%%%%%%%%%%%%%%%%%%%%%%%%%%%%%%%%%%%%%%%%%%%%%%%%%
\subsection{The supersymmetric black ring}
%%%%%%%%%%%%%%%%%%%%%%%%%%%%%%%%%%%%%%%%%%%%%%%%%%%%%%%%%%%%%%%%%%%%%%%%%%%%%%%%%%%%%%%%%

The supersymmetric black ring with three equal charges and three
equal dipole moments has a gravitational field given by
\cite{ElvEmpMatReall1}
\begin{eqnarray}
& & ds^2=-f^2\left (dt+ \omega_{\psi} d \psi+ \omega_{\phi} d \phi
\right )^2+ f^{-1} ds^2(\mathbb{R}^4), \nonumber \\
& & ds^2(\mathbb{R}^4)=\frac{R^2}{(x-y)^2} {\biggl [}
\frac{dy^2}{y^2-1}+(y^2-1)d\psi^2 \nonumber \\
& & \hspace{3cm} +\frac{dx^2}{1-x^2}+(1-x^2)d\phi^2 {\biggr ]},
 \label{black ring}
\end{eqnarray}
where the range of coordinates is $-1\leq x \leq 1$, $-\infty \leq
y \leq -1$. The infinity is at $x=y=-1$, and the event horizon is
at $y=-\infty$. The period of the angular coordinates are $\Delta
\phi=2\pi$ and $\Delta \psi=2\pi$. The constant $R$ sets the
radius of the black ring. The functions appearing in
 (\ref{black ring}) are
\begin{eqnarray}
& & f^{-1}= 1+\frac{Q-q^2}{2R^2}(x-y)-\frac{q^2}{4 R^2}(x^2-y^2), \nonumber \\
& & \omega_{\psi}=\frac{3}{2}q(1+y)+\frac{q}{8R^2}(1-y^2)\left [
3Q-q^2(3+x+y) \right ], \nonumber \\
& & \omega_{\phi}=
 -\frac{q}{8R^2}(1-x^2)\left [ 3Q-q^2(3+x+y) \right ].
 \label{black ring functions}
\end{eqnarray}
The electromagnetic potential of the supersymmetric black ring is
given by
\begin{eqnarray}
 A&=&\frac{\sqrt{3}}{2} {\biggl [} f(dt+\omega_{\psi} d\psi+\omega_{\phi} d\phi)
 \nonumber \\
& & \hspace{0.5cm} -\frac{q}{2} \left [ (1+x)d\phi +(1+y)d\psi
\right ] {\biggr ]}.
 \label{black ring Maxwell}
\end{eqnarray}
The constants $Q$ and $q$ are proportional to the total electric
charge and to the dipole charge, respectively. One has $-1\leq x
\leq 1$, $-\infty \leq y \leq -1$, so we demand that $Q\geq q^2$
in order to guarantee that $f^{-1}\geq 0$ and that the metric has
the correct signature. It is also possible to show that in order
to avoid naked closed timelike curves we must demand that
$R<(Q-q^2)/(2q)$. If this condition is satisfied the solution can
also be extended through $y=-\infty$, and thus this hypersurface
is a regular event horizon. Note also that $\omega_{\phi}(x=\pm
1)=0$ and $\omega_{\psi}(y=-1)=0$ which insures that there are no
Dirac-Misner string pathologies. For a detailed account of the
properties of this solution see
\cite{ElvEmpMatReall1,ElvEmpMatReall2}.

% The mass, angular momenta along $\psi$ and $\phi$, the total
% electric charge, and the dipole charge are respectively given by
% \begin{eqnarray}
% & & M=\frac{3\pi}{4G}\,Q \,, \nonumber \\
% & & J_{\phi}=\frac{\pi}{8G}\,q(3Q-q^2)\,, \nonumber \\
% & & J_{\psi}=\frac{\pi}{8G}\,q(6R^2+3Q-q^2)\,, \nonumber \\
% & & Q_{ADM}=\frac{2}{\sqrt{3}}\, M \,,\nonumber \\
% & & D=\frac{\sqrt{3}}{16 G}\, q\,.
%  \label{black ring functions2}
% \end{eqnarray}

%%%%%%%%%%%%%%%%%%%%%%%%%%%%%%%%%%%%%%%%%%%%%%%%%%%%%%%%%%%%%%%%%%%%%%%%%%%%%%%%%%%%%%%%
\subsection{The wave equation of the supersymmetric black ring}
%%%%%%%%%%%%%%%%%%%%%%%%%%%%%%%%%%%%%%%%%%%%%%%%%%%%%%%%%%%%%%%%%%%%%%%%%%%%%%%%%%%%%%%%%
We now consider the evolution of a minimal scalar in the geometry
(\ref{black ring}). The BPS ring solution of minimal 5D SUGRA has no
scalars built in, but we can consider generic embeddings into 10D
string theory or 11D M theory, thereby getting minimal scalars in
the reduced general 5D theory (corresponding for example to a
graviton polarized along the internal dimensions
\cite{maldacenastrominger}). The evolution of a minimal scalar field
$\Phi$ is governed by the curved space Klein-Gordon equation
(\ref{klein}). We have tried several ansatz, but the one that yields
the simplest expression is again the ansatz (\ref{ansatz}), this
time with both $m_{\phi}$ and $m_{\psi}$ given simply by an integer
number since the periods of the angular coordinates of the
supersymmetric black ring are $\Delta \phi=2\pi$ and $\Delta
\psi=2\pi$. We get the following wave equation
 \beq
  & & 0=(x^2-1)(x-y)^2(y^2-1)f^3 {\biggl [} \frac{\partial}{\partial x}\left ((x^2-1)\frac{\partial
\Psi}{\partial x}\right ) \nonumber \\
& & -\frac{\partial}{\partial y}\left ((y^2-1)\frac{\partial
\Psi}{\partial y}\right ){\biggr ]}+\Psi  {\Biggl [}\omega ^2R^2(x^2-1)(y^2-1) \nonumber \\
& &+ (x-y)^2f^3 {\bigl [} -m
_{\psi}^2(x^2-1)+m_{\phi}^2(y^2-1)\nonumber \\
& &+ (y^2-1)\omega _{\phi}(2m_{\phi}\omega + \omega ^2\omega
_{\phi}) \nonumber \\
& & + (x^2-1)\omega _{\psi}(2m_{\psi}-\omega \omega _{\psi})
{\bigr ]} {\Biggr ]}
 \eeq
This wave equation cannot be separated, at least in these
coordinates. However, as occurred with the non-supersymmetric case,
in the infinite-radius limit it is possible to do the separation of
the wave equation. We turn our attention to this case in the next
subsections.

%%%%%%%%%%%%%%%%%%%%%%%%%%%%%%%%%%%%%%%%%%%%%%%%%%%%%%%%%%%%%%%%%%%%%%%%%%%%%%%%%%%%%%%%%
\subsection{The supersymmetric black string}
%%%%%%%%%%%%%%%%%%%%%%%%%%%%%%%%%%%%%%%%%%%%%%%%%%%%%%%%%%%%%%%%%%%%%%%%%%%%%%%%%%%%%%%%%

The infinite-radius limit of the five-dimensional supersymmetric
black ring yields the supersymmetric black string first found in
\cite{Bena}. To obtain this limit \cite{ElvEmpMatReall1}, one
defines a charge density
\begin{eqnarray}
\bar{Q}=\frac{Q}{2R},
 \label{charge density}
\end{eqnarray}
and the new coordinates
\begin{eqnarray}
r=-\frac{R}{y}, \qquad x=\cos\theta, \qquad \eta=R\psi \,.
 \label{black string coord}
\end{eqnarray}
Taking the $R\rightarrow \infty$ limit, while keeping $\bar{Q}$,
$q$, $r$ and $\eta$ fixed, one gets the gravitational field of the
supersymmetric black string. In this infinite radius limit of a
(supersymmetric) black ring, the charge of the ring becomes a charge
density of the black string, and the dipole charge of the ring
becomes a conserved charged of the string. The black string geometry
is described by
\begin{eqnarray}
 ds^2&=&-f^2\left (dt+ \omega_{\eta} d \eta \right )^2 \nonumber \\
& &  + f^{-1}\left [d\eta^2+dr^2+ r^2(d\theta^2+\sin^2\theta
d\phi^2) \right ] \label{black string}
\end{eqnarray}
where
\begin{eqnarray}
& & f^{-1}= 1+\frac{\bar{Q}}{r}+\frac{q^2}{4 r^2}\,, \nonumber \\
& & \omega_{\eta}=-\left [
\frac{3q}{2r}+\frac{3q\bar{Q}}{4r^2}+\frac{q^3}{8 r^3} \right ],
\label{black string functions}
\end{eqnarray}
and the electromagnetic potential is given by
\begin{eqnarray}
 A=\frac{\sqrt{3}}{2} \left [ f(dt+\omega_{\eta} d\eta)
 -\frac{q}{2} \left ( (1+\cos \theta)d\phi -\frac{d\eta}{r} \right ) \right
 ]\!\!\!.
 \label{black string Maxwell}
\end{eqnarray}

From the discussion of last subsection and using
 (\ref{black string coord}), one knows that the event horizon of the
solution is at $r=0$. One has
$g_{\eta\eta}=(3/q^2)(\bar{Q}^2-q^2)+\mathcal{O}(r)$, thus in
order to avoid closed timelike curves near the horizon (this
condition also guarantees absence of closed timelike curves in the
full solution) we must demand that \cite{GauntGut1}
\begin{eqnarray}
\bar{Q}^2> q^2.
 \label{black string param range}
\end{eqnarray}
Moreover, $g_{t\eta}=2r/q+\mathcal{O}(r^2)$ and therefore the
angular velocity of the event horizon is
\begin{eqnarray}
 \Omega_H=-\frac{g_{t\eta}}{g_{\eta\eta}}{\biggl |}_{r=0}=0.
 \label{black string veloc}
\end{eqnarray}
A supersymmetric asymptotically flat black object, must have a
non-rotating event horizon (this feature was first identified in
the case of the BMPV black hole \cite{GauntMyersTown}). The
supersymmetric black string has a translational invariance along
the string direction and is flat in the transverse directions, and
its horizon is also not rotating. In particular, this means that
there is no ergosphere, and thus we cannot remove kinetic energy
from the supersymmetric black string using the Penrose process or
a superradiant extraction phenomena.

The area per unit length  of the supersymmetric black string
horizon is
 \be
 {\cal A}_H= \pi q \sqrt{3(\bar{Q}^2- q^2)}\,.
 \label{area sym}
 \ee

%%%%%%%%%%%%%%%%%%%%%%%%%%%%%%%%%%%%%%%%%%%%%%%%%%%%%%%%%%%%%%%%%%%%%%%%%%%%%%%%%%
\subsection{\label{sec:separation super}The wave equation of the supersymmetric black string}
%%%%%%%%%%%%%%%%%%%%%%%%%%%%%%%%%%%%%%%%%%%%%%%%%%%%%%%%%%%%%%%%%%%%%%%%%%%%%%%%%%
The evolution of a scalar field $\Phi$ is again dictated by the
Klein-Gordon equation (\ref{klein}), and this time we try the
ansatz:
 \be \Phi=\frac{\Psi(r)}{r} e^{-{\rm i}({\omega t}-m\eta)}
\,,\label{ansatz3}
 \ee
which yields the following radial wave equation
 \be \frac{\partial
^2\Psi}{\partial r^2}+\left ( \frac{\omega ^2}{f^3}-(m+\omega
\omega _{\eta})^2\right )\Psi=0\,.
 \label{radial eq supersym}
\ee

In the next subsection we will find the reflection coefficients
and the absorption cross-section of a scalar wave impinging upon a
supersymmetric black string from infinity. We will use again the
method of matched asymptotic expansions, in which we divide the
space outside the event horizon in two regions, namely, the
near-region, and the far-region, and then we will match the two
solutions in the intersection region.

%%%%%%%%%%%%%%%%%%%%%%%%%%%%%%%%%%%%%%%%%%%%%%%%%%%%%%%%%%%%%%%%%%%%%%%%%%%%%%%%%%
\subsection{\label{sec:absorption supersym}The absorption cross-section of the supersymmetric black string}
%%%%%%%%%%%%%%%%%%%%%%%%%%%%%%%%%%%%%%%%%%%%%%%%%%%%%%%%%%%%%%%%%%%%%%%%%%%%%%%%%%

For simplicity we will consider only the case $m=0$. In the near
region, i.e., in the vicinity of the horizon $r=0$, the radial
wave equation (\ref{radial eq supersym}) can be written as
 \be
  \frac{\partial
^2\Psi}{\partial r^2}+\left
(\frac{3q^2(\bar{Q}^2-q^2)}{16r^4}\omega ^2 \right
)\Psi=0\,.\label{nearr}
 \ee
 The general solution of this equation is
\be
 \Psi=A\,r\,e^{-i\frac{\sqrt{a}}{r}\omega}
+B\,r\,e^{i\frac{\sqrt{a}}{r}\omega} \,,\ee
 where
\be a=\frac{3q^2(\bar{Q}^2-q^2)}{16}\,,\ee
 is a positive real number as a consequence of
 condition (\ref{black string param range}).
 We want only ingoing waves at the horizon so we impose $A=0$.
We also set (this is just a normalization) $B=1$. For small $a$,
the large-$r$ behavior of this solution is
 \be
 \Psi \sim r+i\sqrt{a}\omega\,. \label{near sol sym}
 \ee

On the other hand, in the far region, i.e., for $r \rightarrow
\infty$, the radial wave equation (\ref{radial eq supersym})
reduces to the wave equation in the flat background:
 \be
  \frac{\partial
^2\Psi}{\partial r^2}+\omega ^2\Psi=0\,,\label{farr}
 \ee
which has the plane wave solutions,
 \be
 \Psi = Ce^{i\omega r}+De^{-i\omega
r}\,. \ee
 For small $r$ this solution behaves as
\be \Psi \sim C(1+i\omega r)+D(1-i\omega r)\,. \label{far sol sym}
 \ee

We are now able to  perform a match between (\ref{near sol sym})
and (\ref{far sol sym}), yielding
 \be
C=i\frac{-1+\sqrt{a}\omega
^2}{2\omega}\,,\,\,\,D=i\frac{1+\sqrt{a}\omega ^2}{2\omega}\,.\ee
For small $\sqrt{a}\omega^2$, we have
 \be
  \frac{|C|^2}{|D|^2} \sim 1-4\sqrt{a}\,\omega ^2\,.
 \ee
Thus, the reflection coefficient ${\cal R}\equiv
1-\frac{|C|^2}{|D|^2}$ is given by
\be {\cal R}= q\,\sqrt{3(\bar{Q}^2-q^2)}\,\,\omega ^2\,.\label{rc}
\ee
We have also computed numerically this reflection coefficient, and
the results agreed with (\ref{rc}) to typically within better than
$10^{-2}\,\%$ for small values of $\omega\,,\,q\,,\,\bar{Q}$
(small here means that $\omega \bar{Q}^2$ and $\omega q^2$ are
less than $10^{-3}$).

We can also compute the relevant fluxes for the absorption
cross-section. The conserved flux associated to the radial wave
equation (\ref{radial eq supersym}) is
\begin{eqnarray}
 \mathcal{F}= \frac{2\pi}{i} \left( \Psi^*\partial_r \Psi
 -\Psi \partial_r \Psi^* \right)
  \,.
 \label{flux sym}
\end{eqnarray}
The incident wave from infinity is $\Psi_{\rm inc}=De^{-i\omega
r}$, and thus the incident flux from infinity is given
 by
\begin{eqnarray}
\mathcal{F}_{\rm inc}= 4\pi |D|^2 \omega\,.
 \label{flux inc sym}
\end{eqnarray}
On the other hand the ingoing wave across the horizon is $\Psi_{\rm
abs}=r\,e^{i\frac{\sqrt{a}}{r}\omega}$, and therefore the flux
absorbed by the black string horizon is given
 by
 \begin{eqnarray}
 \mathcal{F}_{\rm abs}= 4\pi \sqrt{a}\, \omega\,.
 \label{flux abs sym}
\end{eqnarray}
The absorption cross-section is given by $\sigma_{\rm abs}
=(\pi/\omega^2) \mathcal{F}_{\rm abs}/\mathcal{F}_{\rm inc}$,
where the factor $\pi/\omega^2$ converts the partial wave
cross-sections into plane wave cross-sections. Explicitly, the
absorption cross-section is then given by
 \begin{eqnarray}
 \sigma_{\rm abs}= \frac{4\pi \sqrt{a}}{1+2 \sqrt{a}\omega^2}
 \sim {\cal A}_H \,,
 \label{cross section sym}
\end{eqnarray}
where in the last approximation we have assumed small values of
$\sqrt{a}\omega^2$, and the fact that ${\cal A}_H=4\pi \sqrt{a}$.
Thus the absorption cross-section is equal to the area, as occurs
generally with spherical black holes \cite{DasGibbonsMathur}.

As a check to (\ref{rc}) and (\ref{cross section sym}), we note
that the reflection coefficient ${\cal R}$ can be written as
${\cal R}= (\mathcal{F}_{\rm inc}-\mathcal{F}_{\rm out})
/\mathcal{F}_{\rm inc}=\mathcal{F}_{\rm abs}/\mathcal{F}_{\rm
inc}$. So one has $\sigma_{\rm abs}=(\pi /\omega^2){\cal R}$, as
it should be.

%%%%%%%%%%%%%%%%%%%%%%%%%%%%%%%%%%%%%%%%%%%%%%%%%%%%%%%%%%%%
\section{Conclusions}
%%%%%%%%%%%%%%%%%%%%%%%%%%%%%%%%%%%%%%%%%%%%%%%%%%%%%%%%%%%%%%%%%%%%
In \cite{hawkingross,prestidge}, the authors have shown how to
separate the wave equation for massless fields in the C-metric
background. Motivated by the fact that the non-supersymmetric black
ring solution \cite{EmpReall} can be constructed by Wick-rotating
the electric charged Kaluza-Klein C-metric, we have tried to
separate the wave equation in the black ring background using the
same ansatz. Even though the wave equation simplifies considerably,
it is still not separable. A similar problem exists in the
Hamilton-Jacobi equation for geodesics in this spacetime
\cite{emparanprivate}. While we cannot rule out separability, it is
likely that an investigation of waves in this geometry will have to
be done using full 2-D numerical simulations.

The only instance where we were able to separate the wave equation
in the full geometry was for time-independent perturbations. A study
of the static equation indicates that one can anchor a scalar field
to a rotating black ring, and this seems to indicate that a black
ring can have scalar hair, at least in this perturbative approach.

On the other hand, the problem of wave propagation simplifies
considerably in the infinite-radius limit of the
non-supersymmetric black ring (which yields a boosted black
string), where the equation separates. We computed the absorption
cross section of the infinite-radius non-supersymmetric black
ring, and we show it is not simply given by the area of the
horizon; it also depends on the frequency of the wave and on the
height $k$ of the potential barrier at infinity. We have shown
that this geometry is stable against scalar perturbations, using
both an analytical and a numerical approach. In principle, one
might still expect the Gregory-Laflamme instability \cite{gl} to
be at work here, and so the geometry might be unstable to
gravitational perturbations.

We have also studied propagation of scalar waves in the background
of a supersymmetric black ring \cite{ElvEmpMatReall1}. Again, in the
infinite radius limit that yields the supersymmetric black string of
\cite{Bena}, we have been able to separate the wave equation. The
scalar low energy absorption cross-section is given by the area per
unit length of the black string horizon. This value should be
reproduced directly by use of correlation functions, once the
correct microscopic description of the five-dimensional BPS black
rings is understood.

%\vskip3.5cm
%%%%%%%%%%%%%%%%%%%%%%%%%%%%%%%%%%%%%%%%%%%%%%%%%%%%%%%%%%%%%%%%%%%%
\section*{Acknowledgements}
%%%%%%%%%%%%%%%%%%%%%%%%%%%%%%%%%%%%%%%%%%%%%%%%%%%%%%%%%%%%%%%%%%%%
We warmly thank Roberto Emparan for useful and enlightening
correspondence on a number of points, and  Henriette Elvang, David
Mateos and Robert Myers for a careful reading of the manuscript. V.
C. and O.J.C.D. acknowledge financial support from Funda\c c\~ao
para a Ci\^encia e Tecnologia (FCT) - Portugal through grant
SFRH/BPD/2004. O.J.C.D. also acknowledges CENTRA - Centro
Multidisciplinar de Astrof\'{\i}sica, Portugal for hospitality. S.Y.
is supported by the Grant-in-Aid for the 21st Century COE ``Holistic
Research and Education Center for Physics of Self-organization
Systems'' from the ministry of Education, Science, Sports,
Technology, and Culture of Japan. This work was partly supported by
the Grant-in-Aid for Scientific Research from the ministry of
Education, Science, Sports, Technology, and Culture of Japan
(17740155).
%%%%%%%%%%%%%%%%%%%%%%%%%%%%%%%%%%%%%%%%%%%%%%%%%%%%%%%%%%%%%%%%%%%%

%%%%%%%%%%%%%%%%%%%%%%%%%%%%%%%%%%%%%%%%%%%%%%%%%%%%%%%%%%%%%%%%%%%%%%%%%%%%%%%%%%%%%%%%%%%%%%
\appendix
\section{\label{sec:A0}Time-independent perturbations in the non-supersymmetric black ring}
%%%%%%%%%%%%%%%%%%%%%%%%%%%%%%%%%%%%%%%%%%%%%%%%%%%%%%%%%%%%%%%%%%%%%%%%%%%%%%%%%%%%%%%%%%%%%%
An interesting particular situation where the wave equation in the
full geometry separates is the static case. For time-independent
perturbations, i.e., $\omega=0$, the wave equation (\ref{kg}) for
the non-supersymmetric black ring can be separable. Indeed, in this
case (\ref{kg1}) yields
\beq & &\!\!\!\!\!\!0= \frac{\partial}{\partial x}\left
(F(x)G(x)\frac{\partial}{\partial x} \Psi \right )-
\frac{\partial}{\partial y}\left (F(y)G(y)\frac{\partial}{\partial
y} \Psi \right )+
\nonumber \\
& &\!\!\!\!\!\! \Psi \left [ \frac{F(y)^2}{G(y)}m_{\psi}
^2-\frac{F(x)^2}{G(x)}m_{\phi} ^2+\nu \left (x-y-2\lambda
(x^2-y^2)\right ) \right ]\,. \nonumber \\ \label{kgf2} \eeq
which is a separable equation. Using the ansatz $\Psi=X(x)Y(y)$ we
get
\beq & & \frac{\partial}{\partial x}\left
(F(x)G(x)\frac{\partial}{\partial x} X \right )+\nonumber \\ & & X
\left [-\frac{F(x)^2}{G(x)}m_{\phi} ^2+\nu x-2\lambda \nu x^2\right
]=LX\, \label{sepx} \eeq
and
\beq & & \frac{\partial}{\partial y}\left
(F(y)G(y)\frac{\partial}{\partial y} Y \right )+\nonumber
\\ & & Y \left [-\frac{F(y)^2}{G(y)}m_{\psi} ^2+\nu y-2\lambda \nu
y^2\right ]=-LY\, \label{sepy} \eeq
where $L$ is a separation constant.

%%%%%%%%%%%%%%%%%%%%%%%%%%%%%%%%%%%%%%%%%%%%%%%%%%%%%%%%%%%%%%%%%%%%%%%%%%%
\subsection{The $x$ equation}
%%%%%%%%%%%%%%%%%%%%%%%%%%%%%%%%%%%%%%%%%%%%%%%%%%%%%%%%%%%%%%%%%%%%%%%%%%%
Near $x=1$, and for $m_{\phi} \neq 0$ equation (\ref{sepx})
behaves as \be X \sim (x-1)^{\alpha}\,,\ee where \be
\alpha=\frac{1}{2} \left ( 1 \pm \sqrt{1+\frac{m_{\phi}
^2(1-\lambda)}{(1-\nu)^2}} \right )\,.\label{eigx1}\ee
 For a black ring in equilibrium satisfying (\ref{equil cond}) this yields
 \be
\alpha=\frac{1}{2} \left ( 1 \pm \sqrt{1+\frac{m_{\phi} ^2}{1+\nu
^2}} \right )\,.\label{eigx2}\ee
 Near $x=-1$ and $m_{\phi} \neq 0$
we have \be X \sim (x+1)^{\beta}\,,\ee where \be \beta=\frac{1}{2}
\left ( 1 \pm \sqrt{1+\frac{m_{\phi} ^2(1+\lambda)}{(1+\nu)^2}}
\right )\,.\label{eigx3}\ee For $m_{\phi}=0$, we have \be X \sim
A_1+A_2\log{(x \pm 1)}\,\,,\,\,x\rightarrow \pm
1\,.\label{eigx4}\ee

Since the field must be regular everywhere outside the horizon,
one must choose the plus sign equations
(\ref{eigx1})-(\ref{eigx3}), and put $A_2=0$ in (\ref{eigx4}).
Therefore, the equation for $X$ is a standard eigenvalue problem,
which determines the value of the separation constant $L$.
%%%%%%%%%%%%%%%%%%%%%%%%%%%%%%%%%%%%%%%%%%%%%%%%%%%%%%%%%%%%
\subsection{The $y$ equation}
%%%%%%%%%%%%%%%%%%%%%%%%%%%%%%%%%%%%%%%%%%%%%%%%%%%%%%%%%%%%
Near $y=-1$, the wavefunction $Y$ behaves as equations
(\ref{eigx3})-(\ref{eigx4}), with the replacement $m_{\phi}
\rightarrow m_{\psi}$.

Near the horizon $y=1/\nu$, equation (\ref{sepy}) behaves as \be Y
\sim (y-1/\nu)^{\alpha}\,,\ee where \be \alpha=\frac{1}{2} \left ( 1
\pm \sqrt{1+\frac{4m_{\psi} ^2(1-\lambda/\nu)}{(\nu-1/\nu)^2}}
\right )\,.\ee For equilibrium black rings this last expression is
equal to \be \alpha=\frac{1}{2} \left ( 1 \pm
\sqrt{1+\frac{4m_{\psi} ^2\nu^2}{\nu ^4-1}} \right )\,.\ee Using
(\ref{quantm}) this yields \be \alpha=\frac{1}{2} \left ( 1 \pm
\sqrt{1+\frac{4n ^2\nu^2}{\nu ^2-1}} \right )\,,\ee where $n$ is an
integer. Now, for $1+\frac{4n ^2\nu^2}{\nu ^2-1}<0$, which always
happens for $\nu$ sufficiently close to $1$, the square root is a
pure complex number, and any behavior is acceptable. If a scalar
field is well behaved at spatial infinity, it will be well behaved
at the horizon. Thus, it looks like one can anchor a scalar field to
a rotating black ring, and this seems to indicate that a black ring
can have scalar hair, at least in this perturbative approach. Notice
that this s only possible for $\nu$ close to $1$, and thus rotating
Myers-Perry black holes are not included in this case.
%%%%%%%%%%%%%%%%%%%%%%%%%%%%%%%%%%%%%%%%%%%%%%%%%%%%%%%%%%%%%%%%%%%%%%%%%%%%
\section{\label{sec:A1}Gamma function relations}
%%%%%%%%%%%%%%%%%%%%%%%%%%%%%%%%%%%%%%%%%%%%%%%%%%%%%%%%%%%%%%%%%%%%%%%%%%%%

In this appendix we derive some gamma functions relations that are
used in the main body of the text.

i) We start with the relations needed to make the transition from
(\ref{factor cross section}) into (\ref{factor cross section 2}).
Using only the gamma function property, $\Gamma(1+x)=x\Gamma(x)$, we
can show that
\begin{eqnarray}
\frac{\Gamma(l+1-i\,2\Upsilon)}{\Gamma(1-i\,2\Upsilon)}=\prod_{\jmath=1}^l
(\jmath- i 2\Upsilon)\,.
 \label{gamma values 1}
\end{eqnarray}
Moreover, using the properties $\Gamma(1+x)=x\Gamma(x)$ and
$\Gamma(ix)\Gamma(1-ix)=- i \pi/\sinh(\pi x)$, and noting that
$\rho= i |\rho|$ yields
\begin{eqnarray}
& & \hspace{-1cm} \Gamma(l+1-\rho)\Gamma(l+1+\rho)\nonumber \\
 & & = \Gamma( i |\rho|) \,\Gamma(1- i |\rho|)\, i |\rho|
\prod_{\jmath=1}^l
(\jmath^2+\rho^2) \nonumber \\
 & & =\frac{\pi |\rho|}{\sinh(\pi |\rho|)}\prod_{\jmath=1}^l
(\jmath^2+|\rho|^2)\,.
 \label{gamma values 2}
\end{eqnarray}

ii) Now, we extract $\delta \rho$ from
 the gamma functions that appear in (\ref{far field-small x aux})
 in order to get (\ref{far field-small x}).

The properties $M(\tilde{a},\tilde{b},\chi=0)=1$ and
$\Gamma(x)\Gamma(1-x)=\pi /\sin(\pi x)$ (with
$x=\tilde{b}-\tilde{a}$ and then with $x=\tilde{b}-1$) allow us to
write (\ref{far field-small x aux}) as
\begin{eqnarray}
\Psi \!\!&\sim&\!\! \beta \, \chi^{l} e^{-\chi/2}\, {\biggl [}
\frac{\sin[\pi (\tilde{b}-\tilde{a})]}{\sin(\pi \tilde{b})}
\frac{\Gamma(\tilde{b}-\tilde{a})}{\Gamma(\tilde{b})}
+\chi^{1-\tilde{b}}\,
\frac{\Gamma(\tilde{b}-1)}{\Gamma(\tilde{a})} {\biggr ]}.\nonumber \\
& &
 \label{far small x A1}
\end{eqnarray}
Use of (\ref{Whittaker parameters end}) with $\delta\rho \sim 0$
yields
\begin{eqnarray}
\frac{\sin[\pi (\tilde{b}-\tilde{a})]}{\sin(\pi \tilde{b})}
\frac{\Gamma(\tilde{b}-\tilde{a})}{\Gamma(\tilde{b})} \sim (-1)^n
\frac{(2l+1+n)!}{(2l+1)!}\,.
 \label{far small x A2}
\end{eqnarray}
To simplify the second term in between brackets in
 (\ref{far small x A1}), we use $\Gamma(\tilde{a})\Gamma(1-\tilde{a})=\pi /\sin(\pi \tilde{a})$
with $\tilde{a}$ defined in (\ref{Whittaker parameters end}) to get
\begin{eqnarray}
& & \frac{1}{\Gamma(-n-i\delta \rho)} \sim -\frac{n!}{\pi}
 \sin[\pi (n+i\delta \rho)] \sim (-1)^{n+1} n! i\delta \rho \,, \nonumber \\
& &
 \label{far small x A3}
\end{eqnarray}
where in the first approximation we used  $\delta \rho \sim 0$ to
obtain $\Gamma(1+n+i\delta \rho) \sim \Gamma(1+n)$, and in the
second approximation we used $\sin(x+y)=\sin x \cos y+ \cos x \sin
y$ together with $\sin(i \pi \delta\rho)\sim i \pi \delta\rho$
(valid for small $\delta \rho$).

The relations  (\ref{far small x A1})-(\ref{far small x A3}) allow
us to go from (\ref{far field-small x aux}) into
 (\ref{far field-small x}).

iii) Finally, we derive the relations needed to make the transition
from (\ref{Matching Instab})
 into (\ref{delta rho}). Use of
$\Gamma(1+x)=x\Gamma(x)$ yields
\begin{eqnarray}
  & & \frac{\Gamma(l+1- i \,2\Upsilon)}
   {\Gamma(-l- i  \,2\Upsilon)}
   = i \,(-1)^{l+1}2\Upsilon \prod_{\jmath=1}^l
(\jmath^2+4\Upsilon^2)\,, \nonumber \\
& & \frac{\Gamma(-2l-1)}{\Gamma(-l)}=(-1)^{l+1}
\frac{l!}{(2l+1)!}\,. \label{gamma values 3}
\end{eqnarray}

%%%%%%%%%%%%%%%%%%%%%%%%%%%%%%%%%%%%%%%%%%%%%%%%%%%%%%%%%%%%%%%%%%%%%%%%%%%%%%%%%%%%%%%%%%%%%%%%%
\section{\label{sec:A2}Numerical approach to the computation of the characteristic frequencies}
%%%%%%%%%%%%%%%%%%%%%%%%%%%%%%%%%%%%%%%%%%%%%%%%%%%%%%%%%%%%%%%%%%%%%%%%%%%%%%%%%%%%%%%%%%%%%%%%%

In this appendix we describe the Frobenius expansion method that is
used to obtain the numerical results on the stability presented in
Sec. \ref{sec:Numerical Instab}. The problem is reduced to the
computation of a continued fraction, which is rather easy to
implement \cite{leaver}.

 The radial equation
(\ref{boost}) can be cast in the same form as that presented in
\cite{NozMaeda}, if we define $\Psi=\frac{1}{r}\Phi$. In this case
$\Phi$ satisfies
\be \frac{d^2}{dr_*^2}\Phi+V\Phi=0\,,\ee
 with $dr/dr_*=1-r_H/r$, and
\beq V &=& f\left ( \frac{\omega
^2}{\bar{f}}-\frac{f'}{r}-\frac{l(l+1)}{r^2}\right ) \nonumber
\\
& & -\bar{f} \left ( k+\frac{r_H\sinh
2\sigma}{2r\bar{f}}\omega\right )^2\,.\label{wave2} \eeq
We can set $r_H=1$ and measure everything in units of $r_H$. The
wave equation (\ref{wave2}) has the asymptotic behavior (ingoing
waves at the horizon)
\be \Phi \sim (r-1)^{-i\left (\omega \cosh \sigma-k \sinh
\sigma\right )}\,\,,\,\,\,\,r\rightarrow 1 \,.\ee Near infinity we
have \be \Phi \sim e^{i\sqrt{\omega ^2-k^2}
\,r_*}\,\,,\,\,\,\,r\rightarrow \infty \,.\ee
The perturbation function $\Phi$ can be expanded around the horizon
as
 \beq\Phi&=&e^{i\omega_\infty
r}r^{i\{\omega_\infty+\omega_H
+(2\omega_\infty)^{-1}(k\cosh\sigma-\omega\sinh\sigma)^2\}}
\nonumber \\& & \times (r-1)^{-i\omega_H}\sum_{k=0}^\infty
a_k\left({r-1\over r}\right)^k\,, \label{expansion}
 \eeq
 where
$a_0$ is taken to be $a_0=1$. Here, $\omega_H$ and $\omega_\infty$
have been defined as
\begin{equation}
\omega_H=\omega\cosh\sigma-k\sinh\sigma\,,\qquad
\omega_\infty=\sqrt{\omega^2-k^2}\,,
\end{equation}
where the branch of $\sqrt{z}$ has been chosen such that
$-\pi/2<{\rm arg}(\sqrt{z})\le\pi/2$. The expansion coefficients
$a_k$ in equation (\ref{expansion}) are determined from the
three-term recurrence relation, given by
\begin{eqnarray}
& & \alpha_0a_1+\beta_0a_0=0\,, \qquad
\alpha_ka_{n+1}+\beta_na_n+\gamma_na_{n-1}=0 \nonumber \\
& & \hspace{3 cm} ({\rm for}\ n=1,2,3,\cdots) \,,  \nonumber
\end{eqnarray}
where
\begin{eqnarray}
& & \hspace{-0.2 cm}
\alpha_n=32(1+n)(1+n+2ik\sinh\sigma-2i\omega\cosh\sigma)
\omega_\infty^2\,, \nonumber \\
& &\hspace{-0.2 cm} \beta_n = \nonumber \\
& & -16(k^2-\omega^2)[-2+(-5+3\cosh{2\sigma})k^2-2l(l+1)\nonumber \\
& & -4n-4n^2+4i\omega\cosh\sigma+8i\omega
n\cosh\sigma+5\omega^2\nonumber \\
& & +3\omega^2\cosh{2\sigma}-
 2ik\{(2+4n)\sinh\sigma-3i\omega\sinh{2\sigma}\}]
\nonumber \\
&&
+8i[(7+\cosh{2\sigma}+14n+2n\cosh{2\sigma}-i\omega\cosh{3\sigma}\nonumber \\
& & - 15i\omega\cosh\sigma)\omega^2
+i(\sinh{3\sigma}-15\sinh\sigma)k^3
+(-7 \nonumber \\
& & -14n+\cosh{2\sigma}+2n\cosh{2\sigma}-
  3i\omega\cosh{3\sigma}
\nonumber \\
& & +15i\omega\cosh\sigma)k^2+\{-2(1+2n)\sinh{2\sigma}+
  3i(\sinh{3\sigma}\nonumber \\
& & +5\sinh\sigma)\omega\}k\omega]\omega_\infty
 \,, \nonumber \\
& & \hspace{-0.2 cm} \gamma_n=\nonumber
\\ & & (-35+28\cosh{2\sigma}-\cosh{4\sigma})\,k^4- 4\{16i n
\sinh\sigma \nonumber\\ &&
+(14\sinh{2\sigma}-\sinh{4\sigma})\omega\}k^3+4k\omega^2\{16i
n\sinh\sigma\nonumber
\\ && +(14\sinh{2\sigma}+\sinh{4\sigma})\omega\}
+\{-32\,n^2+64in\omega\cosh\sigma\nonumber
\\ && +2(35-3\cosh{4\sigma})\omega^2\}k^2
-\{-32\,n^2 + 64i n\omega\cosh\sigma \nonumber \\ &&
+(35+28\cosh{2\sigma}+\cosh{4\sigma})\omega^2\}\omega^2
+8\omega_\infty{\Bigl [}(\sinh{3\sigma}\nonumber
\\ && -7\sinh\sigma)k^3+\{-2i(-3+\cosh{2\sigma})n
-3\omega\cosh{3\sigma}\nonumber \\ &&+7\omega\cosh\sigma\}k^2
-2i(3+\cosh{2\sigma})n\omega^2+\omega^3(\cosh{3\sigma}+\nonumber
\\ && 7\cosh\sigma)\! +\!(4in\sinh{2\sigma}+3\omega\sinh{3\sigma}+7\omega\sinh\sigma)k\omega
{\Bigr ]} \!. \nonumber
\end{eqnarray}

One can see that the expanded eigenfunction $\Phi$ satisfies the
quasinormal boundary conditions at the horizon and at infinity if
the series expansion in equation (\ref{expansion}) converges at
spatial infinity. This convergence condition can be converted into
an algebraic equation for the frequency $\omega$ given by the
continued fraction equation~\cite{leaver,Gu67},
\begin{eqnarray}
\beta_0-\alpha_0\gamma_1 \left (\beta_1-\frac{\alpha_1\gamma_2}
{\beta_2-\frac{\alpha_2\gamma_3} {\beta_3-...}} \right )^{-1} =0
\,. \label{a-eq}
\end{eqnarray}
This continued fraction equation can be solved numerically
\cite{n-recipes}.

%%%%%%%%%%%%%%%%%%%%%%%%%%%%%%%%%%%%%%%%%%%%%%%%%%%%%%%%%%%%%%%%%%%%

\end{document}